\documentclass[twocolumn,useAMS,usenatbib]{mn2e}
\usepackage{graphicx}
\usepackage{bm}

\newcommand{\rmd}{{\rm d}}
\newcommand{\rme}{{\rm e}}

\newcommand{\mrp}{M_{r}}

\title[Intrinsic alignments \& shear calibration]{Galaxy-galaxy weak 
  lensing in SDSS: intrinsic alignments and shear calibration errors}

\author[Hirata et al.]
 {Christopher M. Hirata$^1$\thanks{Electronic address:
    {\tt chirata@princeton.edu}},
  Rachel Mandelbaum$^1$,
  Uro\v s Seljak$^1$,
\newauthor
  Jacek Guzik$^2$,
  Nikhil Padmanabhan$^1$,
  Cullen Blake$^3$,
  Jonathan Brinkmann$^4$,
\newauthor
  Tamas Bud\'{a}vari$^5$,
  Andrew Connolly$^6$,
  Istvan Csabai$^7$,
  Ryan Scranton$^6$,
\newauthor
  and Alexander S. Szalay$^5$
\\$^1$Department of Physics, Jadwin Hall, Princeton University,
      Princeton NJ 08544, USA
\\$^2$Astronomical Observatory, Jagiellonian University, Orla 171, 30-244
      Krak\' ow, Poland
\\$^3$Princeton University Observatory, Princeton University, Princeton NJ
      08544, USA
\\$^4$Apache Point Observatory, 2001 Apache Point Road, Sunspot NM
      88349-0059, USA
\\$^5$Department of Physics and Astronomy, Johns Hopkins University, 3701
      San Martin Drive, Baltimore MD 21218, USA
\\$^6$University of Pittsburgh, Department of Physics and Astronomy, 3941
      O'Hara Street, Pittsburgh PA 15260, USA
\\$^7$Department of Physics, E\"{o}tv\"{o}s University, Budapest, Pf. 32,
      Hungary, H-1518
}

\date{\today}

\begin{document}
\maketitle

\begin{abstract}
Galaxy-galaxy lensing has emerged as a powerful probe of the dark matter
halos of galaxies, but is subject to contamination if intrinsically
aligned satellites of the lens galaxy are used as part of the source
sample.  We present a measurement of this intrinsic shear using 200,747
lens galaxies from the Sloan Digital Sky Survey (SDSS) spectroscopic
sample and a sample of satellites selected using photometric redshifts.  
The mean intrinsic shear at transverse separations of 30--446$h^{-1}$~kpc
is constrained to be $-0.0062<\Delta\gamma<+0.0066$ (99.9 per cent
confidence, including identified systematics), which limits contamination
of the galaxy-galaxy lensing signal to at most $\sim 15$ per cent on
these scales.  We present these limits as a function of transverse
separation and lens luminosity.  We furthermore investigate shear
calibration biases in the SDSS, which can also affect galaxy-galaxy 
lensing, and conclude that the shear amplitude is calibrated to better 
than 18 per cent.  This includes noise-induced calibration biases in the 
ellipticity, which are small for the sample considered here, but which can 
be more important if low signal-to-noise or poorly resolved source 
galaxies are used.
\end{abstract}


\section{Introduction}

Weak gravitational lensing has emerged as a powerful tool for directly
measuring the matter distribution in the universe (e.g.
\citealt{2001PhR...340..291B}; \citealt{2003ARA&A..41..645R}). One of its
applications has been the measurement of the projected matter density of
galaxies (\citealt{1996ApJ...466..623B};
\citealt{1998ApJ...503..531H}; \citealt{2000AJ....120.1198F};
\citealt{2001ApJ...551..643S};  \citealt{2001astro.ph..8013M};
\citealt{2002MNRAS.335..311G}; \citealt{2003MNRAS.340..609H};
\citealt{2003astro.ph.12036S}; \citealt{2004ApJ...606...67H}), groups, and
clusters (\citealt{2001ApJ...548L...5H}; \citealt{1998ApJ...504..636H};
\citealt{2001ApJ...554..881S}).  In these measurements, the observable
quantity is the tangential shear $\gamma_t$ of distant source galaxies
induced by the lens. It is however possible that some of the ``source''
galaxies whose shears are used to measure lensing may in fact be
physically associated with the lens; in this case they may be
intrinsically aligned with the lens, producing a false lensing signal.  
This false signal may be quantified by the ``intrinsic shear''
$\Delta\gamma$, which is the spurious estimate of the tangential shear
that would be obtained by applying a shear estimator to a population of
physically associated galaxies (see Appendix~\ref{app:dgdef} for a more
precise definition).  This intrinsic shear is essentially a type of galaxy
density-shear correlation, i.e. it measures the correlation between the
shears of some galaxies and the positions of others; it is distinct from
the shear-shear correlations that contaminate the gravitational shear
autocorrelation measured by lensing surveys in the field
(\citealt{2000A&A...358...30V}; \citealt{2000MNRAS.318..625B};
\citealt{2001ApJ...552L..85R}; \citealt{2002ApJ...572...55H};
\citealt{2002A&A...393..369V}; \citealt{2003AJ....125.1014J};
\citealt{2003MNRAS.341..100B}).

Several investigators have investigated this potential contaminant of the
lensing signal.  \citet{2001ApJ...555..106L} measured the density-shear
correlation using 12,122 spiral galaxies.  These authors were primarily
interested in the rotation axes of galaxies and their possible use as
tracers of the large-scale tidal field (and hence the cosmic density
perturbations; \citealt{2000ApJ...532L...5L}), and so they measure
position angles rather than shear.  A crude (and model-dependent)
conversion of their measurements into an intrinsic shear suggests
$\Delta\gamma = -0.0037\pm 0.0025$ at separations $r_{3D}=500h^{-1}$~kpc
(see Appendix~\ref{app:pac}).

\citet{2002AJ....124..733B} used 1,819 satellite galaxies selected
spectroscopically from the Two Degree Field (2dF) Galaxy Redshift Survey
\citep{2001MNRAS.328.1039C} with ellipticities measured in the Automatic
Plate Measuring (APM) survey \citep{1990MNRAS.243..692M} to set an upper
limit on the intrinsic shear of the satellite galaxies of
$|\Delta\gamma|\le 0.01$ (at 95 per cent
confidence)\footnote{\citet{2002AJ....124..733B} reported an upper limit
on the ellipticity of 0.02, which we have converted here to a shear.} for
primary-satellite pairs separated by a transverse distance of
$<350h^{-1}$~kpc.  However, this upper limit is not presented as a
function of transverse separation or lens luminosity. Circularization of
the galaxy images due to the APM point-spread function (PSF) may weaken
the upper limit slightly (by $\sim 10$--$20$ per cent;
\citealt{2002AJ....124..733B}) but this does not alter the conclusion that
intrinsic alignment contamination of the galaxy-galaxy lensing signal
observed by \citet{2000AJ....120.1198F} and \citet{2001astro.ph..8013M} is
limited to $\la 20$ per cent.

There have been other observational studies of intrinsic alignments,
including reported detections of higher-order density-shear statistics
\citep{2002ApJ...567L.111L} and shear-shear correlations
(\citealt{2000ApJ...543L.107P}; \citealt{2002MNRAS.333..501B});
unfortunately there is no simple conversion from these measurements into
an intrinsic shear $\Delta\gamma$.

Theoretical work on galaxy intrinsic alignments has been motivated both by
inherent interest and (more recently) by contamination of cosmic shear
surveys for which the intrinsic-alignment parameter of interest is the
shear-shear correlation instead of the density-shear correlation.  A
number of estimates of the shear-shear correlations -- both in $N$-body
simulations and in analytical models -- have been made
(\citealt{2000ApJ...545..561C}; \citealt{2000MNRAS.319..649H};  
\citealt{2000ApJ...532L...5L}; \citealt{2001ApJ...555..106L};
\citealt{2001MNRAS.320L...7C}; \citealt{2001ApJ...559..552C};  
\citealt{2002astro.ph..5212H}; \citealt{2002MNRAS.335L..89J}).  Usually
the galaxy is approximated as a disk aligned perpendicular to the halo
angular momentum vector (for spirals) or as an ellipsoid homologous with
the ellipticity of its halo (for ellipticals), although comparison to
observations \citep{2003astro.ph.10174H} and simulations of galaxy
formation \citep{2002ApJ...576...21V} suggest that this picture is too
simple and overestimates the shear-shear correlations.  Of these authors,
only \citet{2001ApJ...555..106L} and \citet{2002astro.ph..5212H} estimate
the density-shear correlation (and even then only for spirals).  Using the
rough conversion of Appendix~\ref{app:pac}, the
\citet{2001ApJ...555..106L} prediction corresponds to
$\Delta\gamma=-0.004$ for the spiral sources, the $-$ sign indicating
radial alignment.  \citet{2002astro.ph..5212H} present some predictions
for the two-dimensional (projected) density-ellipticity correlation as a
function of angular separation, but do not provide a numerical estimate
$\Delta\gamma(r)$.

In this paper, we use SDSS spectroscopic galaxies (as lenses) and
photometric galaxies (as sources) in a luminosity-dependent study of the
shear around lens galaxies due to intrinsic alignments.  The large number
of galaxies in the SDSS photometric sample allows tighter constraints on
$\Delta\gamma$ than were obtained by \citet{2002AJ....124..733B}.  It also
provides the statistical power to compute these upper limits as a function
of transverse separation and lens luminosity.  To make the interpretation
for weak lensing as simple as possible, we directly compute $\Delta\gamma$
using a lensing estimator (including PSF corrections) rather than the
position-angle statistics of \citet{2001ApJ...555..106L} or the unweighted
moments computed by the APM survey.

Correct calibration of the shear estimator is necessary in order to
interpret either an intrinsic alignment or a weak lensing-induced shear,
particularly if the shear is detected at high signal-to-noise.  Errors in
shear calibration can come from a variety of effects such as incomplete
PSF correction, selection effects, noise-induced biases, uncertainties in
PSF reconstruction, and incomplete knowledge of the ellipticity
distribution of the source population (\citealt{2000ApJ...537..555K};
\citealt{2002AJ....123..583B}; \citealt{2003MNRAS.343..459H}).  Since we
do not detect an intrinsic alignment signal, the only effect of
calibration uncertainty is a minor degradation of our upper limits on
$\Delta\gamma$.  Nevertheless, the calibration results presented here are
of direct interest for ongoing galaxy-galaxy weak lensing studies with the
SDSS; calibration errors at the $\le$16--18 per cent level found here are
potentially important in those cases where the lensing signal is detected
at $\ga 6\sigma$.  The noise-induced calibration biases have not been
explicitly studied previously; while we find that this effect is
insignificant for our sample of relatively bright ($r<21$) galaxies, the
calibration bias can be large for galaxies detected at low signal-to-noise
ratio, especially if they are also not well-resolved.

This paper is organized as follows: Sec.~\ref{sec:lens} includes a
description of the lens catalog, and Sec.~\ref{sec:source} includes
discussion of the source catalog, including details of the apparent shear
measurement, the associated calibration uncertainties, and the photometric
redshifts.  In Sec.~\ref{sec:uncertainty}, we describe the methods used to
compute uncertainties on the shear. Sec.~\ref{sec:results} includes
results for intrinsic and gravitational shear, and the results of tests
for systematic errors affecting these quantities.  We conclude in
Sec.~\ref{sec:discussion} with a discussion of the implications of these
results.  The noise-induced calibration bias derivation is presented in 
Appendix~\ref{app:kn}.

A note about the cosmological model and units used in this paper: we
compute the proper transverse separation of the lens and source using the
physical angular diameter distance $d_A(z_l)$ to the lens.  This distance
is determined assuming a flat $\Lambda$CDM universe with $\Omega_m=0.27$
and $\Omega_\Lambda=0.73$, in accord with cosmological parameter
determinations from the {\slshape Wilkinson Microwave Anisotropy Probe}
\citep{2003ApJS..148..175S}.  We present distances in $h^{-1}$~kpc, where
$h$ is the reduced Hubble parameter: $H_0=100h$~km/s/Mpc.  When
appropriately scaled by $h$, the angular diameter distance is only weakly
sensitive even to extreme variations in cosmology: for a lens at redshift
$z=0.3$ (among the most distant in our sample; see Fig.~\ref{fig:zdist}),
we find that $d_A(z_l)$ is reduced by 7 per cent if we switch from
$\Lambda$CDM to an open universe ($\Omega_m=0.27$, $\Omega_\Lambda=0$),
and by 12 per cent if we switch to an Einstein-de Sitter universe
($\Omega_m=1$, $\Omega_\Lambda=0$).  Given that we see no detection of
intrinsic alignments, the dependence of our results on $\Omega_m$ and
$\Omega_\Lambda$ is negligible for the purpose of constraining
contamination of the weak lensing signal.

\section{Lens catalog}
\label{sec:lens}

The ``lenses'' used in this investigation are obtained from the SDSS main
galaxy spectroscopic sample (\citealt{2000AJ....120.1579Y};  
\citealt{2002AJ....124.1810S}; \citealt{2003AJ....125.2276B}; 
\citealt{2004astro.ph..3325A}).  Our sample
contains 200,747 galaxies within a solid angle of 2,499 square degrees
(SDSS Sample 12; \citealt{2003AJ....125.2276B}). This sample is
approximately flux-limited at Petrosian magnitude $r<17.77$
\citep{2002AJ....124.1810S}.  (The $r$ filter is centered at 622 nm;  
\citealt{1996AJ....111.1748F}).  We have taken only objects at redshift
$z>0.02$; at lower redshifts, the summation of pairs of galaxies out to
$1h^{-1}$~Mpc transverse separation becomes computationally expensive, and
little is gained anyway because of the small inverse critical surface
density\footnote{The inverse critical surface density is defined by
$\Sigma_c^{-1} = 4\pi GD_{OL}D_{LS}/c^2D_{OS}$, where $D_{OL}$, 
$D_{LS}$, and $D_{OS}$ are the angular diameter distances from the 
observer to the lens,
the lens to the source, and the observer to the source.  This quantity is
useful because it packages the geometrical aspects of lensing together;
the shear signal for a given weak lens at fixed physical transverse
separation $r$ scales $\gamma\propto\Sigma_c^{-1}$.}
$\Sigma_c^{-1}$. These spectra have been
processed by an independent spectroscopic pipeline at Princeton
\citep{schlegel}; comparisons with the survey pipeline show that the
failures occur $<1$ per cent of the time. The objects have been
extinction-corrected using the model of \citet{1998ApJ...500..525S}
extrapolated to the SDSS filters using an extinction-to-reddening ratio
$R_V = 3.1$, and $K$-corrected to $z=0.1$ using {\sc kcorrect~v1\_11}
\citep{2003AJ....125.2348B}.

\begin{figure} 
\includegraphics[width=3in,angle=-90]{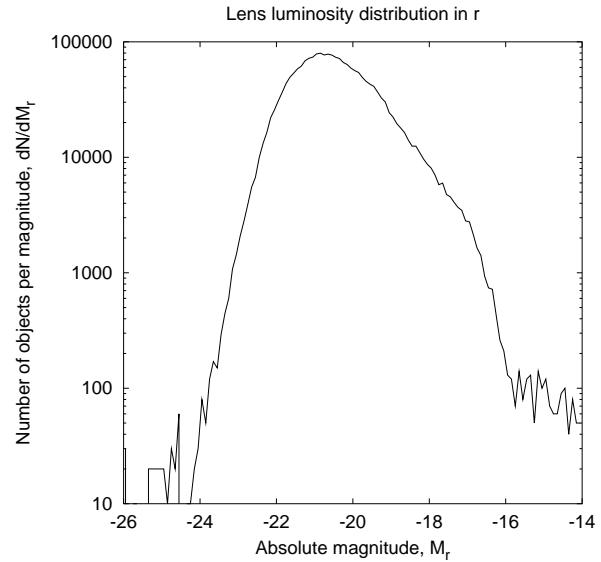} 
\caption{\label{fig:magdist}The magnitude distribution of lens galaxies.}
\end{figure}

\begin{figure} 
\includegraphics[width=3in,angle=-90]{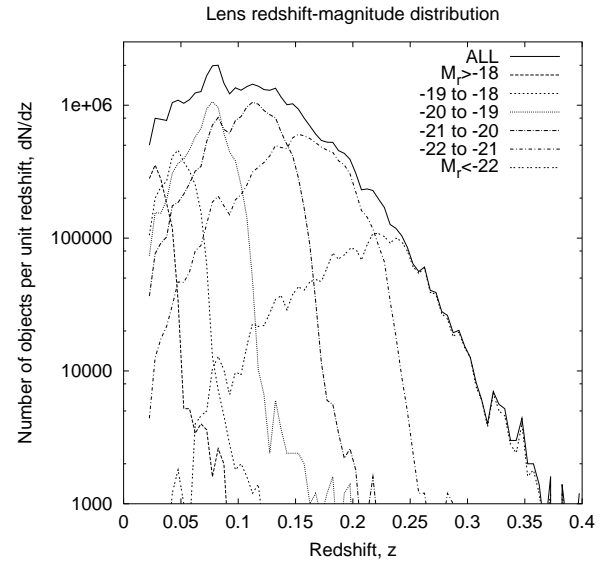} 
\caption{\label{fig:zdist}The redshift distribution of lens galaxies, 
overall and as a function of luminosity.  Note that the brighter (smaller 
$\mrp$) subsamples peak at higher redshift.}
\end{figure}

Fig.~\ref{fig:magdist} shows the magnitude distribution of the lens
galaxies, and Fig.~\ref{fig:zdist} shows their redshift distribution,
overall and as a function of luminosity.  We have used model magnitudes
(i.e. magnitudes determined from the best-fit de Vacouleurs or exponential
profile; \citealt{2002AJ....123..485S}) for both the lens and the source
galaxies in this paper.


\section{Source catalog}\label{sec:source}

\subsection{Apparent shear measurement}

In this section we discuss the methodology for measuring the 
apparent shear of source galaxies.  In general, this must consist of the 
following steps:
\begin{enumerate}
\item Source galaxy detection and selection;
\item Point-spread function (PSF) determination;
\item Measurement of galaxy ellipticity, including PSF correction;
\item Conversion of ellipticity measurement into a shear.
\end{enumerate}
If we intend to measure the radial profile of galaxy halos, we must also
use redshift and cosmological information to convert the shear
measurements into constraints on the mass distribution.  However, in this
paper we are only interested in determining the spurious contribution to
the apparent shear from intrinsic alignment of source and lens galaxies.  
Therefore we only work through the four steps outlined above.  Also, the
reconstruction of the PSF from stellar images is carried out by standard
SDSS software (\citealt{2002AJ....123..485S}; \citealt{lupton}) and will 
not be described here.

\subsubsection{Adaptive moments}
\label{sec:adaptive}

We begin by defining the adaptive moments, which are used in the source
selection and ellipticity measurement. The adaptive second moment matrix
${\mathbfss M}$ of a galaxy's image intensity $I(x,y)$ is found by
minimizing the energy functional:
\begin{equation}
E(A,x_0,y_0,{\mathbfss M};I) = {1\over 2}
\int \left| I(x,y) - A\rme^{-\rho^2/2} \right|^2  \rmd x \rmd y,
\label{eq:energy}
\end{equation}
where:
\begin{equation}
\rho^2 = ({\bmath r}-{\bmath r}_0)^T{\mathbfss M}^{-1}({\bmath r}-{\bmath r}_0)
\label{eq:rho}
\end{equation}
and ${\bmath r}=(x,y)$.  The vector ${\bmath r}_0$ is the object centroid, 
and the moment matrix ${\mathbfss M}$ is taken to be symmetric and can be 
parameterized as:
\begin{equation}
{\mathbfss M} = {T\over 2} \left( \begin{array}{cc} 1+e_+ & e_\times \\ 
e_\times & 1-e_+ \end{array} \right).
\end{equation}
Here $T$ is called the trace, and ${\mathbfss e}=(e_+,e_\times)$
form the spin-2 ellipticity. As defined here the ellipticity is restricted
to lie in the unit circle $e_+^2+e_\times^2\le 1$.\footnote{We warn the
reader that the definition of ellipticity is not standard across the
literature. This definition is consistent with
\citet{2002AJ....123..583B}, but not with \citet{1995ApJ...449..460K}.}
The trace is one measure of the ``size'' of an object; another is the
geometric mean $\sigma$ of the semi major and semi minor axes:
\begin{equation}
\sigma^2 = {T\over 2} \sqrt{1-e^2} = \sqrt{\det{\mathbfss M}},
\end{equation}
where $e^2\equiv e_+^2+e_\times^2$.  The energy functional is minimized by 
requiring:
\begin{equation}
{\mathbfss M} = 2 \frac{ \int ({\bmath r}-{\bmath r}_0)({\bmath 
r}-{\bmath r}_0)^T I(x,y) e^{-\rho^2/2} \rmd x \rmd y
}{ \int I(x,y) e^{-\rho^2/2} \rmd x \rmd y };
\label{eq:emin}
\end{equation}
if the centroid were also determined adaptively, we would simultaneously 
solve
\begin{equation}
0 = \int ({\bmath r}-{\bmath r}_0)I(x,y) e^{-\rho^2/2} \rmd x \rmd y.
\label{eq:emin0}
\end{equation}
However the SDSS photometric pipeline ({\sc photo}) fixes ${\bmath r}_0$ 
at the object finder's centroid  rather than iteratively solving 
Eq.~(\ref{eq:emin0}).  We also use the radial fourth moment $a_4$ defined 
by:
\begin{equation}
2(1+a_4) = \frac{ \int \rho^4 I(x,y) e^{-\rho^2/2} \rmd x \rmd y
}{
\int I(x,y) e^{-\rho^2/2} \rmd x \rmd y
}.
\label{eq:a4}
\end{equation}
{\sc Photo} computes the adaptive moments ${\mathbfss M}$ and $a_4$ for
both the observed (i.e. not deconvolved) galaxy image $I(x,y)$ and the
reconstructed PSF $P(x,y)$.  We will denote moments corresponding to the
galaxy image with a superscript $^{(I)}$ and those corresponding to the
PSF with a superscript $^{(P)}$.  The (raw) resolution factor $R_2$ is
defined by:
\begin{equation}
R_2 = 1 - \frac{ T^{(P)} }{ T^{(I)} };
\end{equation}
in the limit of a very well-resolved galaxy, $R_2\rightarrow 1$, whereas 
for a poorly resolved galaxy the observed image is very similar to the PSF 
($I\approx P$) and so $R_2\rightarrow 0$.

The physical interpretation of ${\mathbfss M}^{(I)}$ is as a best-fit
Gaussian to the image profile.  The fourth moment $a_4^{(I)}$
parameterizes the departure of the galaxy from Gaussianity: it is 0 for a
Gaussian profile, $0.17$ for an exponential profile, and $0.40$ for a de
Vacouleurs profile.  The fourth moment of the PSF diagnoses the deviation
of the PSF from Gaussianity.  In particular, a PSF dominated by Kolmogorov
turbulence (optical transfer function $\propto \rme^{-(l/l_0)^{5/3}}$; 
\citealt{2000ApJ...537..555K}) should have $a_4^{(P)}\approx 0.046$, in 
good agreement with the typical PSF in the SDSS.

\subsubsection{Source galaxy selection}

Our source galaxies are selected from the SDSS photometric catalog
(\citealt{2000AJ....120.1579Y}; \citealt{2001AJ....122.2129H};
\citealt{2002AJ....123..485S}; \citealt{2002AJ....123.2121S};
\citealt{2003AJ....125.1559P}; \citealt{2003AJ....126.2081A}). The catalog
is based on images from the SDSS camera \citep{1998AJ....116.3040G}
processed at Princeton by the {\sc photo} software
(\citealt{2001adass..10..269L}; \citealt{dfink};  \citealt{lupton}). In
order to avoid the PSF anisotropy-induced selection bias discussed by
\citet{2000ApJ...537..555K} and \citet{2002AJ....123..583B}, {\sc photo}
applies a convolution to each image to circularize the effective PSF
before running the object detection algorithm.  We only consider objects
that are classified by {\sc photo} as galaxies, are not deblended, do not
contain saturated pixels, and do not have flags set indicating possible
problems with the measurement of the image. We reject objects whose
adaptive moment measurements failed, the resolution factor
$R_2<\frac{1}{3}$, the measured ellipticity $e_+^{(I)2} +
e_\times^{(I)2}>0.95$, or the radial fourth-moments $|a_4^{(I)}|>0.99$ or
$|a_4^{(P)}|>0.99$ in at least two of the $g$, $r$, and $i$ bands.
Objects with extinction-corrected model magnitude fainter than 21 in
$r$-band or with failed photometric redshifts (see \S\ref{sec:photoz})
are also rejected.  In case of multiple observations of the same object,
we take the observation where the $i$-band resolution factor is largest.

Our final source galaxy catalog contains $N_s=$6,975,528 galaxies,
although some of these are in regions of the SDSS without spectroscopic
coverage and hence are not actually used in the analysis.

\subsubsection{Measurement of galaxy ellipticities}
\label{sec:shapes}

Since most galaxies used in weak lensing measurements are of comparable
angular size to the PSF, the ellipticity $(e_+^{(I)},e_\times^{(I)})$
determined from the galaxy image $I(x,y)$ is generally not equal to the
true ellipticity of the galaxy.  Therefore a correction for the PSF is
necessary.  The basic idea of any such correction is to treat the observed
galaxy image $I$ as a convolution of the PSF $P$ and a pre-seeing image
$f$:
\begin{equation}
I(x,y) = \int f(x',y') P(x-x',y-y') \rmd x' \rmd y'.
\label{eq:fgi}
\end{equation}
The problem is to determine the ellipticity of the pre-seeing image $f$
from $I$ and $P$.  This problem has been considered by many authors
(\citealt{1995ApJ...449..460K}; \citealt{1997ApJ...475...20L};
\citealt{2000ApJ...537..555K}; \citealt{2002AJ....123..583B};
\citealt{2003MNRAS.338...48R}; \citealt{2003MNRAS.343..459H}); we use the
``linear'' method of \citet{2003MNRAS.343..459H} because it can be
implemented using the {\sc photo} outputs.

The ``linear'' method is based on the observation that if the galaxy and
PSF are Gaussian, then the adaptive second moment matrix of $I$ is related 
to that of the PSF and the pre-seeing image by:
\begin{equation}
{\mathbfss M}^{(I)} = {\mathbfss M}^{(f)} + {\mathbfss M}^{(P)},
\label{eq:mf}
\end{equation}
implying:
\begin{equation}
{\mathbfss e}^{(f)} = \frac{ {\mathbfss e}^{(I)} }{ R_2 } 
+ (R_2^{-1}-1) {\mathbfss e}^{(P)}.
\label{eq:mfr2}
\end{equation}
\citet{2003MNRAS.343..459H} made the approximation that the pre-seeing 
galaxy image and the PSF can be approximated by the quartic-Gaussian form:
\begin{equation}
f(x,y) \propto \left[ 1 + {a_4\over 2} (\rho^4 -4\rho^2+2) \right] 
\rme^{-\rho^2/2},
\label{eq:quartic-gaussian}
\end{equation}
where $\rho$ is given by Eq. (\ref{eq:rho}), and similarly for $P(x,y)$.  
Within this approximation, we can compute the corrections to Eq.
(\ref{eq:mf}) to first order in $a_4^{(I)}$ and $a_4^{(P)}$.  The
calculation is straightforward (although tedious) and is given in Appendix
B of \citet{2003MNRAS.343..459H}; it results in a correction to the
resolution factor $R_2$.  We compute ellipticity estimators $(\hat
e_+^{(f)},\hat e_\times^{(f)})$ using the formulas presented there.  
Measurement errors on ${\mathbfss M}^{(I)}$ are computed by {\sc photo}
from the Fisher matrix, and propagated into errors on ${\mathbfss
e}^{(f)}$ via numerical differentiation of the PSF correction formulas.

The actual image of the galaxy and the reconstructed PSF are both
pixelized, and in principle the PSF correction should take this
into account.  Mathematically, pixelization is a two-step process
consisting of convolution with a square top-hat function (i.e.  
integration over the pixel area), followed by sampling at the center of
each pixel.  So long as the convolution step is applied to both the PSF
and the galaxy, it does not cause concern for us because
Eq.~(\ref{eq:fgi}) is still valid.  The sampling at the center of each
pixel replaces the integration in the moment computation with summation
over pixels, which causes a problem for the integrals in
Eq.~(\ref{eq:emin}) if the integrand is insufficiently sampled (i.e. if
there is a high-wavenumber component aliased to zero wavenumber); this
occurs for wavenumbers:
\begin{equation}
l\ge l_{alias}={2\pi\over \theta_{pixel}}\approx 3.28\times 10^6
\approx 15.9$~arcsec$^{-1}.
\end{equation}
Aliasing is more likely to occur for the PSF measurement than the galaxy
measurement since the PSF is smaller in real space.  If
the PSF were a Gaussian with $\theta_{FWHM}=1.2$~arcsec (better than
typical seeing in the SDSS), the function $P({\bmath r})  e^{-\rho^2/2}$
(where $\rho^2$ is determined from Eq.~\ref{eq:rho} using ${\mathbfss
M}^{(P)}$) has $1\sigma$ width of 0.36~arcsec in real space and
2.8~arcsec$^{-1}$ in Fourier-space.  Therefore, we do not expect a
significant contribution to the moment integrals at the aliasing wavenumber
$l_{alias}$, and we have not applied any correction for pixelization.

We have determined the camera shear via differentiation of the SDSS
astrometric solution \citep{2003AJ....125.1559P}, and found it to be
$<0.2$ per cent in all three bands ($g,r,i$) used for the ellipticity
measurement in all six camera columns.  Given that the camera shear is
small and is not correlated with the locations of lens galaxies, no
correction for camera shear has been applied.  Our search for a shear
signal around random points (Sec.~\ref{ss:icsys}) produced a null result, 
confirming that camera shear contamination of the intrinsic alignment 
signal is negligible.

The ellipticities ${\mathbfss e}^{(f)}$ are measured separately in the
$g$, $r$, and $i$ bands (the $u$ and $z$ bands typically have lower
signal-to-noise and so are not as useful for ellipticity measurement).  
An overall ellipticity is computed for each source galaxy by performing an
average (weighted by the measurement error) of the ellipticities in the
bands in which ellipticity measurement was successful.  At this stage we
reject objects with ellipticities $e^{(f)2}>2$.

\subsubsection{Ellipticity to shear conversion}

\begin{figure}
\includegraphics[angle=-90,width=3in]{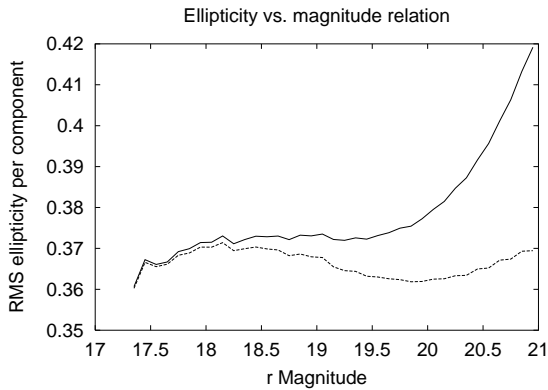}
\caption{\label{fig:e2vmag}
The measured RMS ellipticity of our galaxies per component (i.e. not the
total RMS ellipticity, which is $\sqrt{2}$ times greater since ellipticity
has 2 components) as a function of $r$ magnitude.  The solid curve is the 
RMS of the measured ellipticities, while for the dashed curve the 
measurement noise has been subtracted out.  For brighter galaxies, the
RMS ellipticity per component is $0.37$, which is the value we use to
compute the shear responsivity.  (The statistical significance of the 
downward dip at $r\approx 20$ is large, but its amplitude is only 3 per 
cent, which makes it smaller than possible calibration errors.  It is 
therefore possible that this does not represent a real feature in the 
galaxy ellipticity-magnitude distribution.)}
\end{figure}

Once the ellipticities of the source galaxies have been determined, it is
necessary to convert them into a shear estimator.  The simplest such
estimator is obtained by dividing the ellipticity by an appropriate 
``shear responsivity factor'' ${\cal R}$:
\begin{equation}
\hat\bgamma = \frac{\sum_{j=1}^{N_s} \hat{\mathbfss e}_j}{{\cal R}N_s},
\label{eq:gamma-eu}
\end{equation}
where $N_s$ is the number of source galaxies measured.  Here ${\cal R}$ 
measures the response of the mean ellipticity to an applied shear,
$\langle \hat{\mathbfss e}\rangle = {\cal R}\bgamma + O(\gamma^2)$.  It is 
given by: \citep{2002AJ....123..583B}
\begin{equation}
{\cal R} = 2(1-e_{rms}^2),
\label{eq:response-factor}
\end{equation}
where $e_{rms}$ is the RMS ellipticity per component ($+$ or $\times$).  
From Fig.~\ref{fig:e2vmag}, we find $e_{rms}\approx 0.37$.  (The factor of 
2 comes from the fact that a shear of 1 per cent, when applied to a 
circular object, yields an ellipticity of 2 per cent.)

In principle, better shear signal-to-noise ratio can be obtained if
galaxies are weighted according to the ellipticity measurement error
(\citealt{1997AJ....114...14F}; \citealt{2001astro.ph..8013M}) or if an
ellipticity-dependent weight is used (\citealt{2000ApJ...532...88H};
\citealt{2000ApJ...537..555K}; \citealt{2002AJ....123..583B}).  Since we
only use bright ($r<21$)  galaxies, for which the measurement error is
subdominant to the shape noise, weighting by measurement error is not
helpful.  Also, while ellipticity-dependent weighting is useful for
measurement of gravitational shear, it complicates the interpretation of
intrinsic alignment studies since the apparent shear from intrinsic
alignments probably depends at some level on the method of weighting.  
Therefore in this paper we use only the ``unweighted'' estimator,
Eq.~(\ref{eq:gamma-eu}).

Because there is insufficient signal-to-noise ratio for a shear
measurement around one galaxy, we report the shear measurement averaged in
annular bins around many lens galaxies.  We compute the galaxy-apparent
shear correlation function $\gamma_t(r)$ averaged in the annular bin
$r_-<r<r_+$ by pair summation,
\begin{equation}
\gamma_t(r_-,r_+) = \frac{ \sum_{\alpha,j}\gamma_{t,\alpha j} }{ 
N_p(r_-,r_+) },
\label{eq:gtrr}
\end{equation}
where the summation is over lenses $\alpha$ and sources $j$;
$N_p(r_-,r_+)$ is the number of lens-source pairs with separations 
between $r_-$ and $r_+$; and $\gamma_{t,\alpha j}$ is the $+$ component of 
the apparent shear of source $j$ in coordinates where the $x$-axis is 
aligned perpendicular to the line connecting the lens and source, and the 
$y$-axis is aligned along the line connecting the lens and source.  That 
is, $\gamma_t$ is the ``tangential'' shear component (so that 
lensing-induced shears are positive).  In addition to $\gamma_t$, we 
report the shear $\gamma_{45}$ at a 45 degree angle to the lens (the 
$\times$ component in the coordinate system aligned with the lens-source 
direction).  The latter must vanish in the mean by symmetry, but is 
nevertheless useful for confirmation of the uncertainty estimates.

\subsection{Calibration uncertainty}
\label{ss:calib}

In order to interpret the results of either intrinsic or gravitational
lensing shear measurements, we must understand two types of systematic
errors in the shear measurements: additive biases, which are independent
of the shear signal; and multiplicative or calibration biases, which
affect the response of the shear estimator to an actual shear.  Additive
biases (due, e.g. to PSF ellipticity) constitute a source of spurious
power in the shear power spectrum and hence are a serious issue for cosmic
shear surveys that aim to measure this power spectrum.  However, since we
are essentially computing a cross-correlation function between the galaxy
density and shear, the additive bias can be determined by repeating the
galaxy-shear correlation function measurement using a random ``lens''
galaxy catalog (see Sec.~\ref{ss:icsys}).  Computing the calibration bias
is much harder since there is no straightforward way to measure it
directly.  At present, then, the calibration bias must be understood
theoretically by considering all known sources of calibration error.  The
major sources of error are shown in Table \ref{tab:calib} along with their
estimated uncertainty.  Note that these systematic error estimates are in
some cases very rudimentary and should be considered only as rough guides
to the level at which systematics might be affecting our results.

In addition to the sources of error mentioned in the table, there are two
other phenomena in weak lensing surveys that can mimic a calibration
error: source redshift errors and stellar contamination.  An error in the
source redshift distribution results in incorrect determination of the
lensing strength $\Sigma_c^{-1}$ and hence the surface density contrast
$\Delta\Sigma$; however, as our objective here is to measure apparent
shear rather than mass, this effect is not relevant to the intrinsic
alignment estimation.  Stellar contamination of the ``source galaxy''
sample can dilute the shear signal, resulting in an effective calibration
bias in galaxy-galaxy lensing, however this does not affect our measured
intrinsic shear signal because we divide our intrinsic shear signal by the
fraction of our sources that are physically associated with the lens (as
determined by the lens-source correlation function or ``pair ratio;'' see
Sec.~\ref{sec:results}).  Stars are not clustered around the lens galaxies
and so their dilution effect on the observed shear and the physically
associated fraction cancel out in the estimate of $\Delta\gamma$.

In the remainder of this section we discuss each of the biases in
Table~\ref{tab:calib}.  In each case, our aim is to estimate or constrain 
the induced calibration bias $\delta\gamma/\gamma$, defined by:
\begin{equation}
{\delta\gamma\over\gamma} \equiv {1\over 2} \left.\left(
{\partial \langle\hat\gamma_+\rangle \over \partial \gamma_+} +
{\partial \langle\hat\gamma_\times\rangle \over \partial \gamma_\times} 
\right)\right|_{\bgamma=0}  - 1,
\end{equation}
where $\langle\hat\bgamma\rangle$ is the expectation value of
the shear estimator. Note that $\delta\gamma/\gamma=0$ for an unbiased
shear estimator ($\langle\hat\bgamma\rangle=\bgamma$).  If
$\delta\gamma/\gamma$ is positive, then any galaxy-shear correlation
signal will be overestimated;  if $\delta\gamma/\gamma$ is negative, the
signal will be underestimated.  In principle, $\delta\gamma/\gamma$ may
vary with angular position, however in computing the effective calibration
bias for the galaxy-shear correlation we only care about the average value
of $\delta\gamma/\gamma$.

\begin{table}
\caption{\label{tab:calib}Identified sources of calibration error
$\delta\gamma/\gamma$ in the apparent shear measurement, and rough 
upper limits to the level of error they might induce.  (See text for 
caveats.)}
\begin{tabular}{ll}
\hline\hline
Source & Calibration error \\
& (per cent) \\
\hline
PSF dilution correction & $-8$ to $+6$ \\
PSF reconstruction & $\pm 1.9$ \\
Shear selection bias & $\pm 5.0$ \\
Shear responsivity error & $\pm 1.7$ \\
Noise-rectification bias & $-1.1$ to $+1.4$ \\
\hline
Total & $-18$ to $+16$ \\
\hline\hline
\end{tabular}
\end{table}

\subsubsection{PSF dilution correction}

As noted in Sec.~\ref{sec:shapes}, we do not measure the galaxy
ellipticity ${\mathbfss e}^{(f)}$ directly but rather the measured image
ellipticity ${\mathbfss e}^{(I)}$, which has been diluted by the blurring
effect of the PSF.  Our correction for this effect is not perfect since
the quartic-Gaussian form (Eq.~\ref{eq:quartic-gaussian}) is not a perfect
model for the ellipticity of the galaxy or PSF, and this can lead to a
calibration error.

The calibration error in the PSF dilution correction for adaptive moment
methods was studied in detail in \citet{2003MNRAS.343..459H} (an analysis
for the non-adaptive moment methods can be found in
\citealt{2001A&A...366..717E}). \citet{2003MNRAS.343..459H} found that for
the range of parameters of interest here, the ``linear'' correction can
have calibration bias between $-8$ and $+13$ per cent (the worst case 
being
exponential-profile galaxies whose resolution factor is near our limit
$R_2\approx\frac{1}{3}$).  However, we note that for 81 per cent of our source
galaxies, $R_2\ge\frac{1}{2}$, and within this range
\citet{2003MNRAS.343..459H} found that $\delta\gamma/\gamma$ ranges from
$-8$ to $+4$ per cent.  Using this tighter constraint for the
$R_2\ge\frac{1}{2}$ galaxies, and the $-8$ to $+13$ per cent range for the
19 per cent of our sample with $\frac{1}{3}\le R_2\le\frac{1}{2}$, we
constrain $\delta\gamma/\gamma$ for the overall sample to lie in the range
of $-8$ to $+6$ per cent.  This is the range of values shown in
Table~\ref{tab:calib}.

It can be seen from Table~\ref{tab:calib} that the PSF dilution correction
is currently the largest item in the calibration error budget; this error
will be reduced in future studies by using more accurate but
computationally expensive PSF correction methods such as
re-Gaussianization \citep{2003MNRAS.343..459H} or shapelet decomposition 
(\citealt{2003MNRAS.338...35R}; \citealt{2003MNRAS.338...48R}).

\subsubsection{PSF reconstruction}

The SDSS PSF is reconstructed using images of the bright stars by the {\sc
psp} pipeline (\citealt{2002AJ....123..485S}).  Any error in the PSF model
used in the PSF correction can translate into an error in the
``PSF-corrected'' ellipticity ${\mathbfss e}^{(f)}$. Two major types of
error concern us here:  ellipticity errors and trace (size) errors.  
Ellipticity errors introduce spurious power into the shear measurements,
which can be a serious problem for shear autocorrelation measurements
(\citealt{2003AJ....125.1014J};  \citealt{2004MNRAS.347.1337H}); however
to lowest order these are less serious for galaxy-galaxy lensing because
these errors are not expected to correlate with the locations of galaxies.  
If the size $T^{(P)}$ of the PSF is systematically over- or
underestimated, this leads respectively to under- or overestimation of the
ellipticity ${\mathbfss e}^{(f)}$, and hence to a shear calibration error.
We have tested the PSF traces by comparing them to the observed traces
$T^{(I)}$ of ``stars'' (as identified by {\sc photo}) with PSF magnitude
uncertainty\footnote{The PSF magnitude of an object is computed by fitting
a PSF to the image of the object and finding the best-fit normalization.}
in the range $0.05$--$0.1$ (1$\sigma$).  Fainter stars are not used
because their moments are too noisy; brighter stars are not used because
they have been used as part of the PSF fitting and hence the errors in
their moments may not be representative of the typical error in PSF
reconstruction.

An estimate of the systematic error in the PSF trace can be obtained by
considering the quantity $q=\ln(T^{(I)}/T^{(P)})$.  If the PSF
reconstruction were perfect, we would have (to second order in $\delta
T^{(I)}=T^{(I)}-T^{(P)}$):
\begin{equation}
\langle q\rangle = \frac{ \langle \delta T^{(I)} \rangle }{T^{(P)}}
-{1\over 2} \frac{ \langle \delta T^{(I)2} \rangle }{T^{(P)2}}
= {2\over\nu^2},
\end{equation}
where $\nu$ is the signal-to-noise ratio, and we have used
Eq.~(\ref{eq:dqt}) for the bias and variance of the adaptive trace
estimator, assuming the ellipticity of the PSF is small compared to unity.  
The signal-to-noise ratio, for near-Gaussian PSF, is related to the PSF
magnitude uncertainty $\sigma_{PSFmag}$ via $\nu = 0.4\ln
10/\sigma_{PSFmag}$. Thus the log trace ratio $q$ has expectation value
\begin{equation}
\langle q\rangle = 0.0236 \left( {\sigma_{PSFmag}\over 0.1} \right)^2 + 
O(\sigma_{PSFmag}^3),
\label{eq:qmean}
\end{equation}
which varies from $0.0059$ to $0.0236$ over the range of signal-to-noise
ratios considered.

The tail-rejected means obtained from the stars are $\langle
q\rangle=-0.0087$, $-0.0086$, and $-0.0086$, in $g$, $r$, and $i$
respectively if objects more than $3\sigma$ from the mean are rejected;
and $+0.0057$, $+0.0057$, and $+0.0056$ (with rejection at 7$\sigma$).  
Comparison of these numbers to Eq.~(\ref{eq:qmean}) suggest that the
systematic bias in $q=\ln(T^{(I)}/T^{(P)})$ induced by PSF reconstruction
errors is $\le 3$~per cent, i.e.  that if $T^{(P)}$ is being
systematically over- or under-estimated by {\sc psp} then the magnitude of
this effect is no more than 3 per cent. We have therefore computed the
systematic error in Table~\ref{tab:calib} assuming $\delta
T^{(P)}/T^{(P)}=\pm 0.03$.  The conversion to a shear calibration is
obtained by differentiation of Eq.~(\ref{eq:mfr2}) with respect to
$T^{(P)}$, neglecting the non-Gaussian correction to $R_2$.  This yields,
to first order,
\begin{equation}
{\delta\gamma\over\gamma} = {\delta e_+^{(f)}\over e_+^{(f)}}
= -{\delta R_2\over R_2} = (R_2^{-1}-1){\delta T^{(P)}\over T^{(P)}}.
\end{equation}
(The error for $e_\times$ is the same.)
The mean value of $R_2^{-1}-1$ for our sample of galaxies is 0.62; 
multiplying by 0.03 gives the systematic error estimate in 
Table~\ref{tab:calib}.

\subsubsection{Shear selection bias}

If the algorithm for selecting galaxies preferentially selects galaxies
that are nearly circular, then the estimated shear (as computed from
Eq.~\ref{eq:gamma-eu}) will be closer to zero than the true shear.  This
could arise, e.g. because after PSF convolution a circular galaxy is
spread over fewer pixels (and hence will be detected at higher
signal-to-noise) than a highly elongated galaxy with the same magnitude
and pre-seeing area.  The actual selection algorithm used here is more
sophistocated than a simple signal-to-noise cut; we expect that some of
our cuts, e.g. the magnitude cut at $r<21$, will be relatively insensitive
to shear selection biases whereas others, e.g. the resolution factor cut
at $R_2\ge{1\over 3}$, will not.  In principle the shear selection bias
would be eliminated if the selection algorithm consisted of a simple cut
on the magnitude of the galaxy (which is shear-independent), however in
practice one needs additional cuts, e.g. to remove unresolved galaxies
from the catalog.

The shear selection bias is difficult to compute from first principles
because of cuts made on the galaxy sample (convergence of adaptive
moments, flags, resolution factor cut, ellipticity cut, magnitude 
cut, etc.).  Of these,
the most worrying are the resolution factor cut and the adaptive moment
convergence.  In general, the shear selection bias is given by
\citep{2003MNRAS.343..459H}:
\begin{equation}
{\delta\gamma_+\over\gamma_+} = {1\over\cal R} \left\langle e^{(f)}_+
{\partial\over\partial\gamma_+}\ln{\cal P} \right\rangle,
\label{eq:dgg}
\end{equation}
where the average is taken over galaxies used for the measurement; ${\cal
P}$ is the probability of a galaxy being detected and selected; $e$ is the
ellipticity of the galaxy; and $\gamma$ is the gravitational shear. (A
similar equation holds for $\delta\gamma_\times$.) Conceptually, this
equation is telling us that if the selection probability is decreased by
applying an infinitesimal shear that makes the galaxy more elliptical,
then the shear will be underestimated.

\begin{figure}
\includegraphics[angle=-90,width=3in]{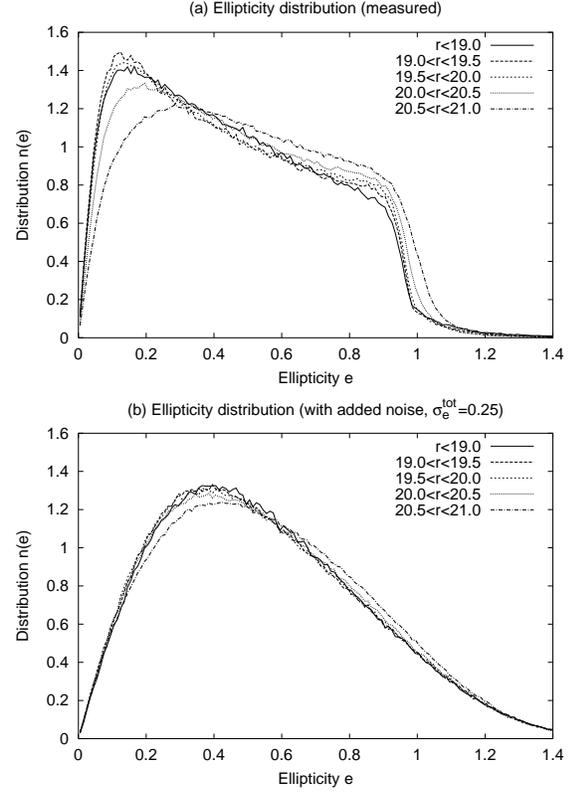}
\caption{\label{fig:ellip}(a) The distribution of measured PSF-corrected
ellipticity $\hat e^{(f)}=\sqrt{e_+^2+e_\times^2}$ of source galaxies as a
function of $r$ band magnitude.  Each curve has been separately
normalized to $\int n(e)de=1$.  (b) The same distribution, except that 
each galaxy has had noise added to both ellipticity components so that 
the total noise is $\sigma_e=0.25$ per component.  This allows us to 
directly compare the various ellipticity distributions, since 
differences among them are not due to variations in $\sigma_e$.  A small 
number (ranging from 0.3 per cent at $r<19$ to 7 per cent at 
$20.5<r<21.0$) of objects had $\sigma_e>0.25$; no noise was added to their 
ellipticities.}
\end{figure}

Given the absence of a good theoretical model of selection biases, we
parameterize the problem by assuming that there is a selection probability
that is a function of the ellipticity, ${\cal P}(e)$, and that we may
approximate $\ln{\cal P}(e)$ using a polynomial of order $2\alpha$:
\begin{equation}
\ln{\cal P}(e) = \sum_{j=0}^{\alpha} c_{2j} e^{2j}.
\end{equation}
(Odd-order terms in $e$ are forbidden by symmetry.)  The distribution of 
galaxy ellipticities is shown in Fig.~\ref{fig:ellip}(a).  In 
Fig.~\ref{fig:ellip}(b), we have added noise to the ellipticities so that 
all the galaxies have the same ellipticity noise $\sigma_e=0.25$ (i.e. 
we have added random increments $\Delta e_+$ and $\Delta e_\times$ of 
variance $\sqrt{0.25^2-\sigma_e^2}$ to each component of the 
ellipticity).  If we assume that the difference between the $r<19.0$ and 
$20.5<r<21.0$ curves in Fig.~\ref{fig:ellip}(b) is entirely due to 
selection biases, then we should have:
\begin{equation}
\ln { n_{20.5<r<21.0}(e) \over n_{r<19.0}(e) } = \sum_{j=0}^{\alpha} 
c_{2j} e^{2j} +{\rm const.}
\end{equation}
A fit to this equation in the range $0<e<1.5$ using $\alpha=4$ (5 
parameters) gives $c_2=-0.0178$, $c_4=-0.281$, $c_6=+1.140$, and 
$c_8=+0.205$.  (We fit beyond the maximum $e=1.414$ used in this paper to 
suppress the characteristic ``ringing'' of polynomial fits.)  Substituting 
this fit into Eq.~(\ref{eq:dgg}), and noting that 
$de/d\gamma_+=2e_+(1-e^2)$, we find that the shear selection bias is:
\begin{eqnarray}
{\delta\gamma\over\gamma}
&=& {1\over \cal R}\int e(1-e^2)n(e) {\rmd\ln {\cal P}\over\rmd e} \rmd e
\nonumber \\
&=& {1\over \cal R}\int_0^{e_{max}} (1-e^2)n(e)
\left(\sum_{j=1}^{\alpha}2j c_{2j} e^{2j}\right) \rmd e.
\label{eq:dgg-shear}
\end{eqnarray}
For the $20.5<r<21.0$ magnitude range, we compute
$\delta\gamma/\gamma=+0.025$.  We find that the computed
$\delta\gamma/\gamma$ remains between 0.02 and 0.03 as we adjust the order
of the fitting polynomial from 4 ($\alpha=2$) to 20 ($\alpha=10$),
indicating that the polynomial fit is not affecting the results.  The
error given in Table~\ref{tab:calib} is 5.0 per cent, which is twice the
error obtained from Eq.~(\ref{eq:dgg-shear}) (to give a more conservative
estimate).  We emphasize that we are placing a constraint on the selection
bias, and that we have {\em not} found evidence that this bias is in fact
present.  In particular, the differences between the curves in
Fig.~\ref{fig:ellip}(b) could also be due to the small number of objects
with large $\sigma_e$ (so that their ellipticity noises could not be
``boosted'' to 0.25); non-Gaussianity of the ellipticity error
distribution; systematic errors in ellipticity measurement (perhaps due to 
PSF effects); a real variation in the source ellipticity distribution with
apparent magnitude (since the brighter and fainter galaxies represent 
distinct populations); or some combination of these effects.

\subsubsection{Shear responsivity error}

Any error in $e_{rms}$ translates into an error in the shear
responsivity ${\cal R}$ via Eq.~(\ref{eq:response-factor}) and hence an 
error in the calibration of Eq.~(\ref{eq:gamma-eu}).  While the
statistical uncertainty in $e_{rms}$ is negligible because we have
millions of source galaxies, the systematic uncertainty must be taken into
account.  An error $\delta e_{rms}$ thus translates into a calibration 
error given by:
\begin{equation}
{\delta\gamma\over\gamma} = -{\delta{\cal R}\over{\cal R}}
= {2e_{rms}\delta e_{rms}\over 1-e_{rms}^2}
\approx 0.86\delta e_{rms}.
\end{equation}
From Fig.~\ref{fig:e2vmag}, we take $\delta e_{rms}=\pm 0.02$ as a 
reasonable error estimate; if the error in $e_{rms}$ is larger than 
this then it must either cancel a magnitude dependence of the rms 
ellipticity or else it must affect objects of all magnitudes similarly 
(which is unlikely because the resolution factor $R_2$ is strongly 
correlated with magnitude).  This leads to the error estimate from shear 
responsivity error of $\delta\gamma/\gamma=\pm 0.017$, as given in 
Table~\ref{tab:calib}.

\begin{figure}
\includegraphics[angle=-90,width=3in]{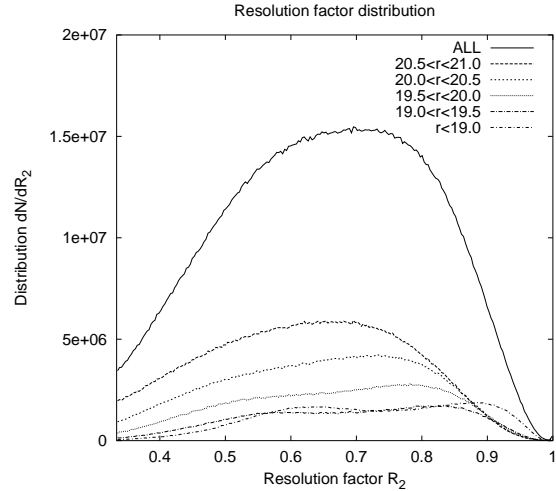}
\caption{\label{fig:rr2}The distribution of source galaxy resolution
factors, $dN/dR_2 = N_sn(R_2)$.}
\end{figure}

\subsubsection{Noise-rectification bias}

The fitting to determine ${\mathbfss M}^{(I)}$, and the subsequent PSF
correction to obtain ${\mathbfss e}^{(f)}$, are nonlinear operations,
hence noise in the original image $I({\bmath r})$ can lead to a bias in
the ellipticity ${\mathbfss e}^{(f)}$.  The additive bias (spurious power)
introduced by noise rectification has been previously noted in the context
of cosmic shear surveys (\citealt{2000ApJ...537..555K};
\citealt{2002AJ....123..583B}). Here we consider the effect of noise
rectification on calibration bias.  This effect has not been previously 
analyzed, although it is present implicitly in simulations
(e.g. \citealt{2001MNRAS.325.1065B}; \citealt{2001A&A...366..717E}).

The calibration correction due to noise-rectification bias is proportional to 
the noise variance and hence to $\nu^{-2}$, where $\nu$ is the detection 
significance of the galaxy, i.e.:
\begin{equation}
{\delta\gamma\over\gamma} \approx K_N\nu^{-2} + O(\nu^{-3}).
\label{eq:nrect}
\end{equation}
The noise-rectification coefficient $K_N$ is evaluated in 
Appendix~\ref{app:kn}, and found to be (compare to 
Eq.~\ref{eq:noiserect}):
\begin{equation}
K_N = 4(1-3R_2^{-1}+R_2^{-2}+2e_{rms}^2).
\end{equation}
Our galaxies have resolution factor ${1\over 3}<R_2<1$ and 
$e_{rms}=0.37$.  Within this range, $K_N$ takes on a minimum value of 
$-3.9$ at $R_2={2\over 3}$ and a maximum value of $+5.1$ at $R_2={1\over 
3}$.  A typical $r=21$ galaxy in moderate (1.7~arcsec FWHM) 
seeing has $\nu\sim 10$, $25$, and $20$ in the $g$, $r$, and $i$ bands, 
respectively.  The bands are weighted by ellipticity error 
$\sigma_e^{-2}$, which is proportional to $\nu^2$ assuming the resolution 
factors in the three bands are all equal.  In this case, the weighted mean 
$\nu^{-2}$ is:
\begin{equation}
\langle\nu^{-2}\rangle_{\rm weighted} = {3\over \nu_{g}^2 + \nu_{r}^2
+ \nu_{i}^2 },
\end{equation}
which for the signal-to-noise ratios listed above case is 0.0027.  The 
product $K_N\nu^{-2}$ appearing in Eq.~\ref{eq:nrect} is thus between 
$-0.011$ and $+0.014$.  This is shown as the upper limit in 
Table~\ref{tab:calib}.

\begin{figure}
\includegraphics[angle=-90,width=3in]{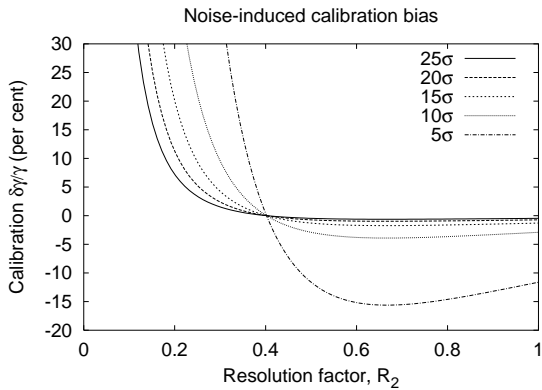}
\caption{\label{fig:knp}The calibration bias obtained from 
Eq.~(\ref{eq:noiserect}) for $e_{rms}=0.37$, as a function of resolution 
factor $R_2$ and detection signal-to-noise ratio $\nu=5...25$.  Note the 
large bias for poor resolution factors $R_2<1/3$ and low signal-to-noise.
The zero crossing at $R_2\approx 0.4$ is somewhat dependent on $e_{rms}$ 
and only exists for the first term in the asymptotic expansion of
Eq.~(\ref{eq:nrect}), hence this plot should not be taken as an 
indication that weak lensing measurements are best done with galaxies of 
this resolution factor.}
\end{figure}

While the noise-induced calibration bias (Eq.~\ref{eq:nrect}) is a
subdominant source of error in this investigation, the same may not be
true of all lensing observations.  For example, for sources detected at
$10\sigma$ ($\nu=10$) and resolution factor $R_2=0.2$, we find
$\delta\gamma/\gamma = +0.44$, which likely dominates the calibration
error budget for these sources (see Fig.~\ref{fig:knp}).  The lensing
studies that use these low-significance objects usually down-weight them
due to their large ellipticity uncertainty $\sigma_e$, so the
noise-induced calibration bias in the final result could still be small.  
However in this situation the ellipticity estimator $\hat{\mathbfss
e}^{(f)}$ can be correlated with the weight, which leads to additional
terms (positive or negative) in the calibration.  The analysis of these
correlations and their induced calibration bias is not needed here and is
deferred to future work.

\subsection{Source photometric redshift samples}
\label{sec:photoz}

It is critical for weak lensing experiments to have accurate knowledge of
the source redshifts (and the lens redshifts, in the case of galaxy-galaxy
lensing studies). It is also useful for studies of the intrinsic
correlations, since it enables us to distinguish physically associated
pairs from widely separated pairs for which the gravitational lensing
effect dominates the galaxy-apparent shear correlation.  In this
investigation, we have used spectroscopic redshifts for the lens galaxies
and photometric redshifts for the source galaxies.  This section describes
the methodology used to determine the photometric redshifts, and briefly
considers some of tests of the photometric redshifts.  Photometric
redshift tests will be discussed in more detail in \citet{mandelbaum}.

\subsubsection{Methodology}

Photometric redshift algorithms using the five SDSS bandpasses ($u$, $g$,
$r$, $i$, and $z$ centered at 354, 475, 622, 763, and 905 nm,
respectively; \citealt{1996AJ....111.1748F}) have previously been
developed \citep{2003AJ....125..580C}.  These photometric redshifts are
based on the ``hybrid'' photometric redshift methods
(\citealt{2000AJ....119...69C}; \citealt{2000AJ....120.1588B}), in which
the spectral energy distribution (SED) templates are parameterized; the
SED parameters and source redshifts are then simultaneously fit to the
photometric data.  These are only available for some regions of sky
(specifically the 2099 deg$^2$ covered by the SDSS Data Release 1,
hereafter DR1).  We have therefore used the following methodology to
derive photometric redshifts: the DR1 galaxies in our sample are placed in
the 5-dimensional $ugriz$-space.  For each source galaxy outside of the
DR1 coverage, we identify the nearest neighbour in $ugriz$-space with a
computed photometric redshift:
\begin{equation}
s = \sqrt{ (\Delta u)^2 + (\Delta g)^2 + (\Delta r)^2 + (\Delta i)^2 
   + (\Delta z)^2 }.
\end{equation}
We report a photometric redshift failure if the nearest neighbour lies at
distance $s>0.1$ in $ugriz$-space.  Of the 8,574,845 galaxies
with measured ellipticities and $r<21$ that pass the flag cuts, matches 
are found for 6,925,528 (including DR1 objects).  The separation $s<0.05$ 
for 5,889,951 of these galaxies, including the 3,046,566 galaxies used to
construct the nearest-neighbour points.

\subsubsection{Pair ratio tests}
\label{ss:pzt}

The pair ratio $R_p(r)$ (Eq.~\ref{eq:d2d}) is a useful tool for testing
photometric redshifts; it is defined for a given transverse separation and
lens luminosity as the ratio of the number density of ``source'' galaxies
in a given redshift slice at that transverse separation to the number
density found in the field. We split the lens-source pairs into widely
separated ($z_s>z_l+\epsilon$, where $\epsilon=0.05$~or~$0.1$) and nearby
($|z_s-z_l|<\epsilon$) samples.  If the photometric redshifts were
perfect, we would have $R_p(r)\approx 1$ for the $z_s>z_l+\epsilon$
samples (the exception to this is the magnification bias effect, which
should be of the same order of magnitude as the shear, i.e. of order 1 per
cent). For the $|z_l-z_s|<\epsilon$ samples, we expect $R_p(r)>1$ due to
galaxy clustering.

\begin{figure} 
\includegraphics[width=3in]{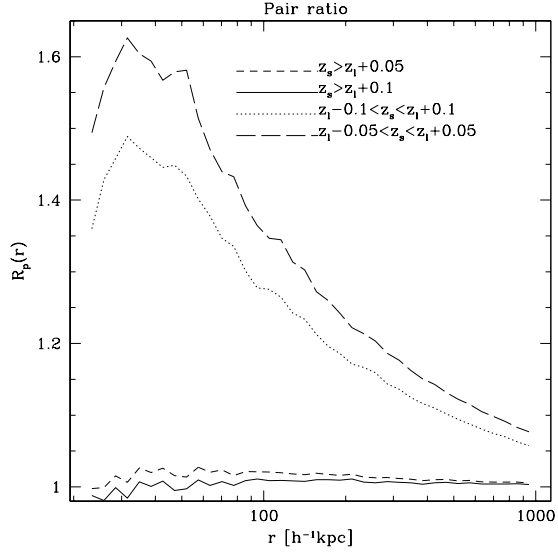} 
\caption{\label{fig:excess}Pair ratio for several samples, as indicated on 
the plot.}
\end{figure}

We compute the pair ratio by dividing the number of lens-source pairs per
unit area by the value obtained from randomly generated (Poisson) lens
catalogs:
\begin{equation}
R_p(r) = \frac{n(r)}{n_{rand}(r)},
\label{eq:fexcess}
\end{equation}
where $n(r)$ is the number of lens-source pairs per unit area in a given
radial bin. The large $R_p$ for the $|z_l-z_s|<\epsilon$ samples is easily
seen in Fig.~\ref{fig:excess}, and it is greatly suppressed in the
$z_s>z_l+\epsilon$ samples. The density of source galaxies varies
considerably with seeing because of our resolution factor and
signal-to-noise cuts.  If the lens selection algorithm were perfectly
independent of seeing, this would not contaminate the pair ratio (which is
essentially a cross-correlation between lenses and sources), but slight
variations in the lens number density with seeing could cause a systematic
effect (positive or negative) on $R_p(r)$. However, in the widely
separated samples the combined effect of any seeing systematics and
lens-source clustering (due to photometric redshift errors) is
$|R_p(r)-1|<0.03$ in the innermost bins ($r<100 h^{-1}$~kpc) and $<0.01$
in the outer bins ($r\sim 1000 h^{-1}$~kpc).  When we split the lenses
into different magnitude ranges, we find $|R_p(r)-1|<0.03$ in all
magnitude ranges except for $\mrp<-22$, where $R_p$ ranges from 1.2 (in
$30<r<100h^{-1}$~kpc) to 1.04 (in $300<r<992h^{-1}$~kpc).  This excess is
more likely to be a result of photometric redshift errors than systematics
in the source density because it is much more significant at small radii
and for the brighter lenses.  This suggests that the seeing systematic in
$R_p(r)$ is at the 1--3 per cent level or less.

\begin{figure} 
\includegraphics[width=3in]{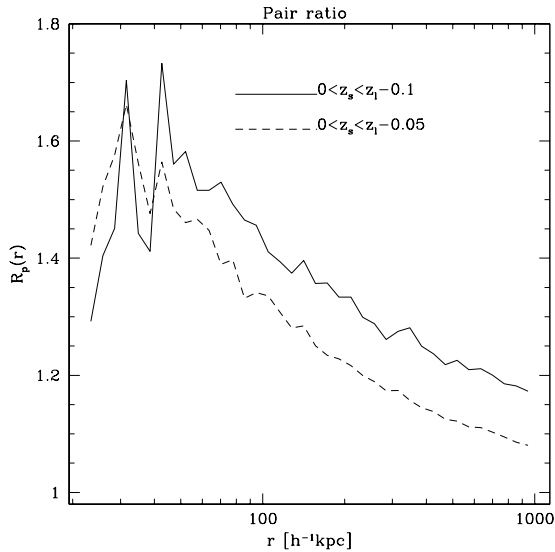} 
\caption{\label{fig:excess.below}Pair ratio for $0<z_s<z_l-\epsilon$.  
Counterintuitively, the $\epsilon=0.1$ curve is above the $\epsilon=0.05$ 
curve; this is because higher-redshift (and hence more luminous due to 
the apparent magnitude cut for SDSS spectroscopy) lenses are weighted more 
heavily when we examine only pairs with the ``source'' at lower redshift.}
\end{figure}

As another test of the photometric redshifts, Fig.~\ref{fig:excess.below}
shows a plot of the pair ratio for galaxies at $0<z_s<z_l-\epsilon$.  
This sample should also include no physically-associated pairs, and
therefore should have $R_p$ near 1.  Instead, there is a large pair ratio
at small radii, indicating that some physically-associated galaxies at
higher redshift are erroneously assigned to very low redshift.  (We
examined the pair ratio in the range $0.003<z_s<z_l-\epsilon$ to eliminate
objects assigned to very low photometric redshift; this did not reduce the
pair ratio, and so we have not imposed this cut in any of the results
presented in this paper.)  Fortunately, however, the actual number of 
source galaxies represented by the excess pair ratio in 
Fig.~\ref{fig:excess.below} is relatively small.  To see this, we compute 
the excess number of galaxies in the $0<z_s<z_l-\epsilon$, 
$z_l-\epsilon<z_s<z_l+\epsilon$, and $z_s>z_l+\epsilon$ samples:
\begin{eqnarray}
N_+ &=& (N - N_{rand})[z_s>z_l+\epsilon] \nonumber \\
N_0 &=& (N - N_{rand})[z_l-\epsilon<z_s<z_l+\epsilon] \nonumber \\
N_- &=& (N - N_{rand})[z_s<z_l-\epsilon],
\end{eqnarray}
and compute the normalized fractions:
\begin{equation}
f_{+,0,-} = \frac{ N_{+,0,-} }{ N_++N_0+N_- }.
\end{equation}
The fractions $f_+$, $f_0$, and $f_-$ respectively represent the fraction 
of physically associated source galaxies whose photometric redshifts are 
overestimated, correctly estimated (to within $\pm\epsilon$), and 
underestimated.  For the sample of all galaxies at radii 
$30<r<446h^{-1}$~kpc, we find $f_+=0.13$, $f_0=0.82$, and $f_-=0.05$ 
(for $\epsilon=0.1$), i.e. only 5 per cent of the source galaxies that are 
clustered with the lens have redshifts underestimated by more than $0.1$.  
The overestimation and underestimation failure rates increase to 
$f_+=0.26$ and $f_-=0.16$ if we narrow our redshift slice to 
$\epsilon=0.05$.  We find that $f_-$ is an increasing function of lens 
luminosity, which is expected since (i) the photometric redshift can be an 
underestimate of the true redshift only if the true redshift exceeds 
$\epsilon$, and (ii) the more luminous lenses tend to be more distant.  
Another trend that we find is that the ``probability of correct source
photometric redshifts'' $f_0$ decreases with increasing 
transverse separation $r$ (e.g. for the all lenses, $\epsilon=0.1$ sample 
we have $f_0=0.90$ at $30<r<100h^{-1}$~kpc, decreasing to $f_0=0.80$ at 
$300<r<446h^{-1}$~kpc), which could be indicative of a variation of 
satellite galaxy colors as a function of radius.  The result that 
$f_+>f_-$ indicates a bias or skewness in the photometric redshift error 
distribution.

It thus appears that for the sources used in this paper, roughly 40 per
cent of the source galaxies physically associated with a lens are
scattered outside of $z_l\pm 0.05$ by photometric redshifts errors, and 20
per cent are scattered outside of $z_l\pm 0.1$.

\subsection{Excluded lens-source pairs}

At sufficiently small lens-source separation angles, the measurement of
sources becomes complicated because the source can overlap with the lens 
(or with the tail of the PSF convolved with the lens).  Problems that 
occur in this regime can include incorrect ellipticity measurement if 
overlapping objects are used in the analysis, and selection bias if they 
are not.  In our case deblended objects are not used for ellipticity 
measurement.

\begin{figure} 
\includegraphics[width=3in]{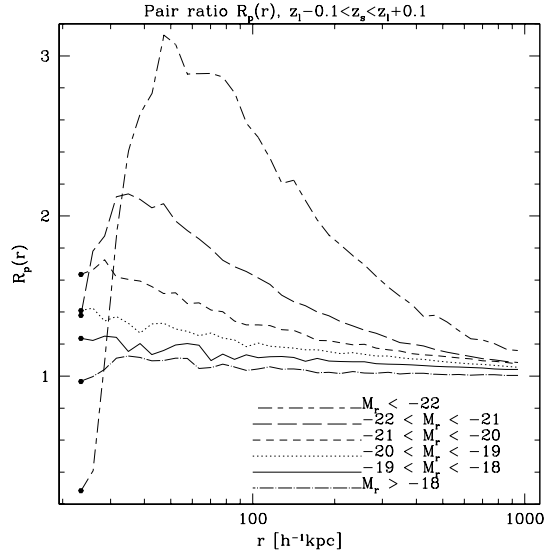} 
\caption{\label{fig:excess.mag}Pair ratio for $|z_s-z_l|<0.1$, several 
luminosity subsamples.}
\end{figure}

A crude way to determine which annular bins are affected by the
overlapping source-lens effect is to examine the pair ratio for source
samples selected to be within redshift $|z_s-z_l|<\epsilon$ of the lens
(Fig.~\ref{fig:excess.mag} for $\epsilon=0.1$).  The effect of clustering
is easily seen: the pair ratio is greater than unity and (over most of the
range of radii) increases as the lens magnitude $\mrp$ increases and the
transverse separation $r$ decreases.  However, at the innermost radii
there is a drop off in $R_p(r)$ since sources inside these radii are
deblended. Brighter galaxies have a more pronounced dip in the excess at
the very innermost radii (out to 40 physical $h^{-1}$~kpc for the very
brightest bin), for two reasons: (i) these bright galaxies are typically
quite large; and (ii) they are more distant so the same angular separation
(relevant to deblending) corresponds to greater physical transverse
separation.  \citet{2004NewA....9..329P} find that for $\mrp=-22$
elliptical galaxies (the fainter edge of this brightest luminosity bin),
the typical 50 per cent light radius is $10h^{-1}$~kpc, in which case
there may be significant overlap at the innermost bin ($r=20h^{-1}$~kpc).
Since the angular diameter distance is $15z_lh^{-1}$~kpc per arc second,
and the typical seeing in the SDSS is 1--2 arc seconds, the deblending
effect can increase the minimum usable radius by tens of $h^{-1}$~kpc for
our brightest subsample (typical $z_l\sim 0.25$).  For the faintest
subsample (typical $z_l\sim 0.03$), this increase is much less important.

Therefore, for the brightest two luminosity samples ($\mrp<-21$),
we have rejected the inner bins where $R_p$ is increasing with $r$.
(We have verified that rejecting one extra bin or one fewer bin does not 
significantly affect the final results.)

%\begin{figure} 
%\includegraphics[width=3in]{excessmag05}
%\caption{\label{fig:excess.mag.05}Pair ratio for $|z_s-z_l|<0.05$, 
%several 
%luminosity subsamples.}
%\end{figure} 

\section{Shear uncertainty estimation}
\label{sec:uncertainty}

We use two methods to compute the statistical uncertainty in the shear
$\gamma_t$.  The first is a random lens method, in which we use our source
catalog to compute the covariance of the shear signal around random points
within the lens sample region.  The second is a bootstrap method in which
we cut the lens sample region into 150 subregions that are then
re-sampled.  Because both methods use the actual source catalog, they will
correctly take into account error from any spurious shear power in the
source catalog.  The random lens method has the disadvantage of not taking
into account fluctuations in the lens number density (particularly if they
correlate with the noise or systematics in the source catalog).  The
bootstrap method addresses these shortcomings on the smallest scales, but
suffers due to the fact that the shear signals computed in the 150
subregions are not truly independent, because a source galaxy near a
subregion boundary may contribute to the measurement of shear around
lenses in multiple subregions.  In addition, the noise in the bootstrap
covariance matrix (which can have a significant impact on, e.g., $\chi^2$
testing) is more difficult to assess.  The bootstrap covariance matrix can
be made essentially noiseless by taking thousands of subregions, but then
the correlations among the subregions become large and the validity of the
bootstrap method is undermined.

Faced with the deficiencies of either the random lens method or the
bootstrap method individually, we have chosen to combine them as follows.
The errors on the intrinsic alignment determination are computed using
both the random-lens and bootstrap covariance matrices; consistency
between the two methods suggests that both are reasonable estimates of the
uncertainty.  The $\chi^2$ analysis is performed using the covariance
matrix from the random lens catalogs; the effect of noise in the
covariance matrix is taken into account as described in
Appendix~\ref{app:chi2}.  The outermost bins in the faintest luminosity
subsamples are excluded from the analysis since galaxy clustering renders
the random-lens method inaccurate here, and it is for these bins that the
independence of subregions assumed by bootstrap has the least robust
justification.

\subsection{Random lens method}

We compute our statistical errors using a random lens test, which involves
computing the shear around randomly-located ``lenses,'' where Poissonian
(independent) random lens locations were generated using the angular mask
of the survey.  We constructed 78 random lens catalogs; the galaxies in
each random catalog were assigned magnitudes and redshifts drawn from the
true sample without replacement.  The covariance matrix between radial
bins $i$ and $j$ is computed as:
\begin{equation}
\hat C_{ij} = {1\over M-1}\sum_{\alpha=1}^{M} \left[
\gamma_t^{(\alpha)}(r_i)
\gamma_t^{(\alpha)}(r_j) - \bar\gamma_t(r_i)\bar\gamma_t(r_j) \right],
\label{eq:sample-cov}
\end{equation}
where $M=78$ is the number of random catalogs, the index $\alpha$ denotes
the random catalogs, and $\bar\gamma_t(r_i)$ is the sample mean of the
shear in radial bin $i$ averaged over all of the random catalogs.  So long
as the error distribution of the $\{\gamma_t(r_i)\}$ is jointly Gaussian,
$\hat{\mathbfss C}$ gives an unbiased estimate of the covariance (but it
cannot be used naively in $\chi^2$ tests; see Appendix~\ref{app:chi2}).  
The number of random catalogs used must be $\ge N$ (the number of radial
bins) in order to produce a non-degenerate estimator $\hat{\mathbfss C}$
for the covariance, but is limited by available processor time.

\begin{figure}
\includegraphics[angle=-90,width=3in]{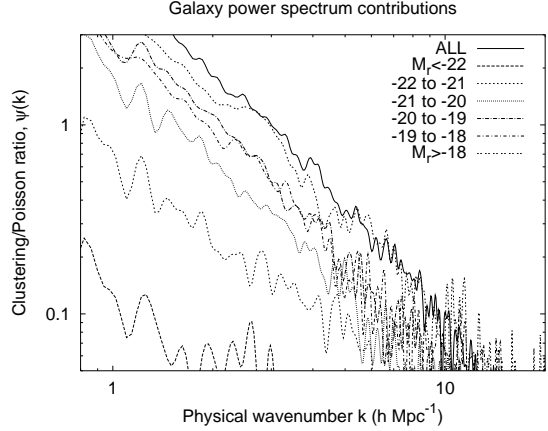}
\caption{\label{fig:auto}The ratio $\psi(k)$ of clustering to Poisson 
contributions to the galaxy power spectrum.  This is obtained from the 
data using Eq.~(\ref{eq:psi}), imposing a smooth cutoff in the integrand 
in the range $20$--$30$~$h^{-1}$~Mpc.}
\end{figure}

The Poisson-distributed random lens method produces valid error bars on
angular scales where the angular (two-dimensional) galaxy power spectrum
is Poisson-dominated and the clustering contribution can be safely
neglected.  Neglecting edge effects in the survey, this range of angular
scales can be determined from the ratio of the galaxy clustering power
spectrum to the Poisson spectrum:
\begin{eqnarray}
\psi(k) &\equiv &\frac{P(k,{\rm clustering})}{P(k,{\rm Poisson})}
\nonumber \\
&=& {2\over N_{gal}} \int_0^\infty {\rmd N_{p,excess}\over \rmd r} J_0(kr) 
\rmd r,
\label{eq:psi}
\end{eqnarray}
where $N_{gal}$ is the number of galaxies in the sample, $k$ is the 
physical ({\em not} comoving) wavenumber of interest, $r$ is the 
transverse separation, and $N_{p,excess}$ is the number of pairs of 
galaxies with transverse separation $\le r$ minus the number of pairs 
derived from a random (Poisson) catalog.  We use the average of the 
angular diameter distances to the two galaxies for the conversion of 
angular separation into transverse separation $r$ (this should be valid 
because no correlations should exist between galaxies at very different 
angular diameter distances).  The ratio $\psi(k)$ is plotted for our 
samples in Fig.~\ref{fig:auto}.

If we were cross-correlating all of the lens galaxies against the same
sample of ``source'' galaxies, the error variance of the cross-spectrum
would simply be re-scaled in proportion to the galaxy power spectrum, i.e.
as $\propto 1+\psi(k)$.  (If survey boundary effects can be neglected it
does not matter that we are computing the correlation function via pair
summation as opposed to via a cross-spectrum in Fourier space.) If
$\psi(k)\ll 1$, then the Poisson power spectrum is a good approximation to
the lens galaxy power spectrum and hence to the error of $\gamma_t$.  If
$\psi(k)\gg 1$, then the lens galaxy power spectrum is much greater than
the Poisson estimate, and the error on $\gamma_t$ for Poisson-distributed
lenses is less than the error for ``real'' clustered galaxies.  Thus we
expect the Poisson random lens method to be a good way to estimate error
bars in the case where $\psi(k)\ll 1$.  In this paper we are
cross-correlating each lens galaxy with the shear of source galaxies
within the redshift slice $z_l\pm\epsilon$; this increases the importance
of the clustering contribution.  However, for each of our luminosity
subsamples, the width of the lens redshift distribution $\Delta z$ is of
order $2\epsilon$ (or less, for the lowest-luminosity subsamples), hence
$\psi(k)\ll 1$ is still a valid condition for the clustering to be
unimportant.  Since the galaxy power spectrum scales roughly as $P(k,{\rm
clustering})\propto k^{-1.2}$, whereas the Poisson contribution is
independent of $k$, the clustering contribution dominates on large scales
whereas the Poisson contribution dominates on small scales.  The random
lens method is thus valid only at the smaller scales.

The above reasoning applies in Fourier space.  The real-space
galaxy-apparent shear correlation function $\gamma_t(r)$ is related to the
galaxy-apparent shear cross-power spectrum by:
\begin{equation}
\gamma_t(r) = {1\over 2\pi} \int_0^\infty k^2 P_{g\gamma}(k)
J_2(kr) \rmd\ln k.
\end{equation}
The $J_2$ function peaks at $kr\approx 3.055$; therefore, we take only the
annuli out to a distance $r_{\rm max} = 3.055/k_{\rm min}$ where
$\psi(k_{\rm min})=0.2$.  The values of $r_{\rm max}$ are shown in
Table~\ref{tab:rmax}.

\begin{table}
\caption{\label{tab:rmax}The number of lens galaxies in each luminosity 
subsample, and the maximum radius $r_{\rm max}$ at which the Poisson
random-catalog uncertainties are valid.  We estimate $r_{\rm max}$ from 
$r_{\rm max}=3.055/k_{\rm min}$, where $k_{\min}$ is the minimum 
wavenumber (largest scale) at which $\psi(k)\le 0.2$, i.e. at which the 
clustering power spectrum is $\le 20$ per cent of the Poisson power 
spectrum.}
\begin{tabular}{ccrcr}
\hline\hline
Subsample & ~ & $N_{gal}$ & ~ & $r_{\rm max}$ ($h^{-1}$~Mpc) \\
\hline
all & ~ & 200747 & ~ & 0.43 \\
$\mrp<-22$ & ~ & 11337 & ~ & $>$1.00 \\
$-22<\mrp\le -21$ & ~ & 54768 & ~ & $>$1.00 \\
$-21<\mrp\le -20$ & ~ & 72464 & ~ & 0.74 \\
$-20<\mrp\le -19$ & ~ & 41276 & ~ & 0.55 \\
$-19<\mrp\le -18$ & ~ & 14475 & ~ & 0.52 \\
$\mrp>-18$ & ~ & 6427 & ~ & 0.43 \\
\hline\hline
\end{tabular}
\end{table}

While the clustering of lens galaxies has the effect of increasing our
error bars, the clustering of source galaxies around lens galaxies tends
to decrease errors because there are more ``sources'' near the lenses in
the real catalog than in the Poisson catalogs.  This effect is significant
when $R_p-1$ is of order unity (or larger), i.e. in the innermost
200~$h^{-1}$~kpc of our brighter subsamples.  It is only important for the
source samples whose redshifts are similar to the lens redshifts (the
$z_l-\epsilon<z_s<z_l+\epsilon$ samples) since there is negligible
lens-source correlation between different redshifts.  We have not applied
any source clustering correction in the computation of our error bars,
thus our error estimates should be viewed as conservative.

\subsection{Bootstrap method}

The bootstrap method, as applied here, consists of the following major 
steps:
\begin{enumerate}
\item Split the lens catalog into $M=150$ subcatalogs containing similar 
numbers of lens galaxies.
\item Generate $K=10^5$ synthesized lens catalogs.  Each synthesized lens 
catalog is a concatenation of $M=150$ of the subcatalogs, selected 
randomly with replacement.  (Some lens galaxies appear in the synthesized 
catalog multiple times, others not at all.)
\item Compute the shear signals 
$\{\gamma^{[\alpha]}_t\}_{\alpha=1}^K$ 
from each of the synthesized catalogs, and compute their covariance 
$\hat{\mathbfss C}_{boot}$.
\end{enumerate}
In order to produce subcatalogs that are as independent as possible, we 
divide galaxies into subcatalogs that consist of different regions of sky.  
The subcatalog boundaries are based on the SDSS $\lambda,\eta$ coordinate 
system \citep{2002AJ....123..485S}.  The lens galaxies are first split 
into 45 rings of width $\Delta\lambda=4$~degrees bounded by parallels of 
constant $\lambda$.  Within each ring, they are then rank-ordered by 
$\eta$ (in the range of $-180$ to $+180$ degrees).  Having arranged the 
galaxies in this order, we cut the catalog into 150 approximately 
equal-sized consecutive groups, which become our subcatalogs.  This 
results in galaxies falling into the same subcatalog as most of their 
neighbours.  (Some subcatalogs consist of multiple ``islands'' in 
different parts of the survey; we have not attempted any detailed 
optimization of the subcatalog cuts.)

The covariance matrix computed via the bootstrap method is noisy even as
$K\rightarrow\infty$ because a finite number $M$ of subcatalogs has been
used.  In the extreme case of $M\le N$, the shear signals from each of the
subcatalogs would form an $M$-dimensional plane in the $N$-dimensional
shear space; then the signals computed from bootstrap re-sampling of these
subcatalogs would all lie in this plane and their covariance matrix would 
be degenerate.  As $M$ increases, the noise in the bootstrap covariance 
matrix goes down, but the samples become less and less independent.  We 
have used $M=150$ here.  Since some of the subcatalogs receive higher 
weight than others (due to survey geometry, variations in source density, 
etc.), the calculation of Appendix~\ref{app:chi2} cannot be directly 
applied to the bootstrap method.  While this makes $\chi^2$ testing using 
the bootstrap method difficult, the errors on the final bin-averaged 
intrinsic alignments obtained from $\hat{\mathbfss C}_{boot}$ are stable 
and provide a good check of the correctness of the random-lens errors.

\subsection{Chi-squared test}

\begin{figure} 
\includegraphics[angle=-90,width=3in]{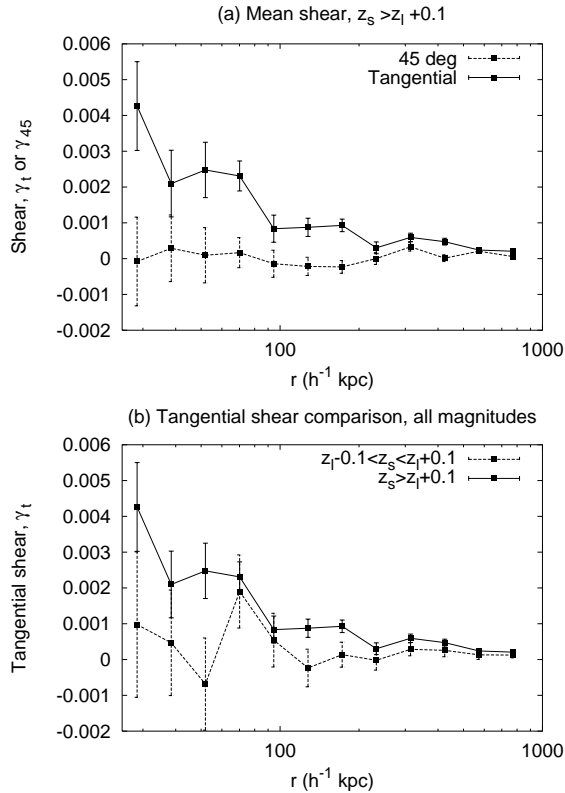} 
\caption{\label{fig:e.and.bmode}(a) Shear signal for $z_s>z_l+0.1$.
(b) Comparison of tangential shear for pairs widely separated vs. nearby 
in redshift.  In both panels we have combined groups of three radial bins 
for clarity.}
\end{figure}

\begin{table}
\caption{\label{tab:chi2}The $\hat\chi^2$ values for fit to zero
  shear (for both the tangential shear $\gamma_t$ and the rotated shear 
  $\gamma_{45}$) for various lens samples.  The $p$ values quoted take 
  into account the effect of noise on the $\chi^2$ calculation as 
  explained in Appendix \ref{app:chi2}, and represent the probability of a 
  random vector having $\hat\chi^2$ less than the given value by chance.  
  The number of usable bins (i.e., radial bins where $r$ is large enough 
  to avoid overlap with the lens galaxy but small enough that the 
  clustering contribution is unimportant) is $N$.}
\begin{tabular}{crrlrl}
\hline\hline
Subsample & $N$ & $\hat\chi^2_{(t)}$ & $p_{(t)}$
& $\hat\chi^2_{(45)}$ & $p_{(45)}$ \\
\hline
\multicolumn{6}{c}{\mbox{$z_s>z_l+0.1$}} \\
\hline
all & 27 & 179.1 & 0.99999 & 70.8 & 0.937 \\
$\mrp>-18$ & 29 & 37.9 & 0.249 & 50.5 & 0.570 \\
$-19<\mrp\le -18$ & 30 & 38.2 & 0.204 & 42.8 & 0.313 \\
$-20<\mrp\le -19$ & 27 & 67.9 & 0.922 & 67.9 & 0.922 \\
$-21<\mrp\le -20$ & 34 & 227.3 & 0.99998 & 65.9 & 0.568 \\
$-22<\mrp\le -21$ & 35 & 160.5 & 0.997 & 57.2 & 0.331 \\
$\mrp<-22$ & 33 & 154.0 & 0.998 & 82.4 & 0.849 \\
\hline
\multicolumn{6}{c}{\mbox{$|z_s-z_l|<0.1$}} \\
\hline
all & 26 & 39.7 & 0.489 & 27.1 & 0.125 \\
$\mrp>-18$ & 26 & 31.6 & 0.244 & 40.7 & 0.517 \\
$-19<\mrp\le -18$ & 28 & 25.5 & 0.051 & 27.9 & 0.081 \\
$-20<\mrp\le -19$ & 30 & 28.3 & 0.044 & 21.9 & 0.007 \\
$-21<\mrp\le -20$ & 32 & 85.4 & 0.901 & 43.9 & 0.225 \\
$-22<\mrp\le -21$ & 33 & 35.5 & 0.060 & 45.1 & 0.201 \\
$\mrp<-22$ & 30 & 35.1 & 0.142 & 32.6 & 0.097 \\
\hline
\multicolumn{6}{c}{\mbox{$z_s>z_l+0.05$}} \\
\hline
all & 27 & 213 & $>0.99999\!\!$ & 75.3 & 0.958 \\
$\mrp>-19$ & 27 & 35.4 & 0.290 & 42.6 & 0.501 \\
$-21<\mrp\le -19$ & 31 & 168.8 & 0.99990 & 88.2 & 0.936 \\
$\mrp<-21$ & 33 & 244 & $>0.99999\!\!$ & 42.7 & 0.157 \\
\hline
\multicolumn{6}{c}{\mbox{$|z_s-z_l|<0.05$}} \\
\hline
all & 26 & 37.0 & 0.411 & 25.1 & 0.085 \\
$\mrp>-19$ & 26 & 34.6 & 0.331 & 14.2 & 0.002 \\
$-21<\mrp\le -19$ & 30 & 35.6 & 0.149 & 28.1 & 0.043 \\
$\mrp<-21$ & 33 & 40.9 & 0.128 & 55.6 & 0.424 \\ \hline\hline
\end{tabular}
\end{table}

We may test the random-catalog errors using a $\chi^2$ test, since the
$\gamma_{45}$ shear must (in the mean) vanish by symmetry. At this point,
it is essential to note that error covariance matrices obtained from
simulations (e.g., as used in this paper) {\em cannot} be blindly used in
place of the true covariance matrices in statistical tests, because they
are obtained from a finite number of simulations and hence have a
statistical error distribution themselves.  In particular, when testing a
signal for consistency with zero, it is common to use the $\chi^2$ test,
based on the variable:
\begin{equation}
\chi^2_{(45)} = {\bmath\gamma}_{45}^T {\mathbfss C}^{-1} 
{\bmath\gamma}_{45}.
\end{equation}
Here ${\bmath\gamma}_{45}=(\gamma_{45}(r_1),...,\gamma_{45}(r_N))$ is the
$N$-dimensional vector of shears in the $N$ radial bins.  If
${\bmath\gamma}_{45}$ is Gaussian-distributed with covariance ${\mathbfss
C}$ and mean ${\bmath 0}$, then $\chi^2_{(45)}$ has a $\chi^2$
distribution with $N$ degrees of freedom.  If we replace ${\mathbfss C}$
with the estimate $\hat{\mathbfss C}$, the distribution of $\chi^2$
becomes much more complicated.  This distribution, while analytically
intractable, is easy to compute via Monte Carlo simulations (see
Appendix~\ref{app:chi2}), and we have used this approach for all the
$p$-values quoted in this paper.

We test for zero $\gamma_{45}$ by computing the statistic
$\hat\chi^2_{(45)}=\bgamma_{(45)}^T{\sf C}^{-1}_{rand}\bgamma_{(45)}$,
where ${\sf C}_{rand}$ is the error covariance matrix derived from random
catalogs. (We put a ``hat'' on the $\hat\chi^2$ as a reminder that the
covariance matrix is noisy, so that this statistic does not have the usual
$\chi^2$ distribution; see Appendix~\ref{app:chi2}.) The shear data are
plotted in Fig.~\ref{fig:e.and.bmode}(a), while the $\hat\chi^2_{(45)}$
values (broken down by lens luminosity) are shown in Table~\ref{tab:chi2}
along with their $p$-values (the cumulative probability of obtaining a
lower $\hat\chi^2$ assuming that there is indeed zero signal, the errors
are Gaussian, and the random realizations correctly sample the error
distribution). Out of the 22 subsamples shown in the table, only 1 has a
$p$-value for the 45-degree shear above 0.95, and 4 have $p$ values above
0.9.  These numbers are consistent with the high $p$-values occurring due
to chance.  However, the high incidence of low $p$-values in the
$|z_s-z_l|<\epsilon$ (two subsamples have $p_{(45)}<0.01$) samples
requires explanation.  This is likely due to conservative assumptions made
in computing the errors, particularly the neglect of source-lens
clustering, which leads to overestimated errors and therefore lower
$\hat\chi^2_{(45)}$ and $p$-values.  Because source-lens clustering is
more significant for the $|z_s-z_l|<\epsilon$ samples than for the
$z_s>z_l+\epsilon$ samples, this explanation is consistent with the fact
that the abnormally low $p$-values occur for the $|z_s-z_l|<\epsilon$
samples.  Thus, we conclude that the $p$-values for the $\gamma_{45}$
shear show no evidence for systematic effects and suggests that the
random-catalog procedure is not detectably underestimating the error bars.

For completeness, we also show in Table~\ref{tab:chi2} the tangential
shear $\hat\chi^2_{(t)}$ and $p$-values.  In this case, however, the
signal is inconsistent with zero for many of the subsamples, with high $p$
for the brighter luminosity lenses and widely separated
($z_s>z_l+\epsilon$)  lens-source pairs.  As can be seen from the
luminosity-averaged shear signal (Fig.~\ref{fig:e.and.bmode}), this is
because the measured $\gamma_t>0$ but only for widely separated pairs,
consistent with gravitational lensing.  Since in this paper we do not fit
a model to the lensing shear signal, the tangential $\hat\chi^2_{(t)}$
values are not directly useful for establishing the accuracy of error
bars.

\section{Results}
\label{sec:results}

\begin{figure*}
\includegraphics[width=6.75in,angle=-90]{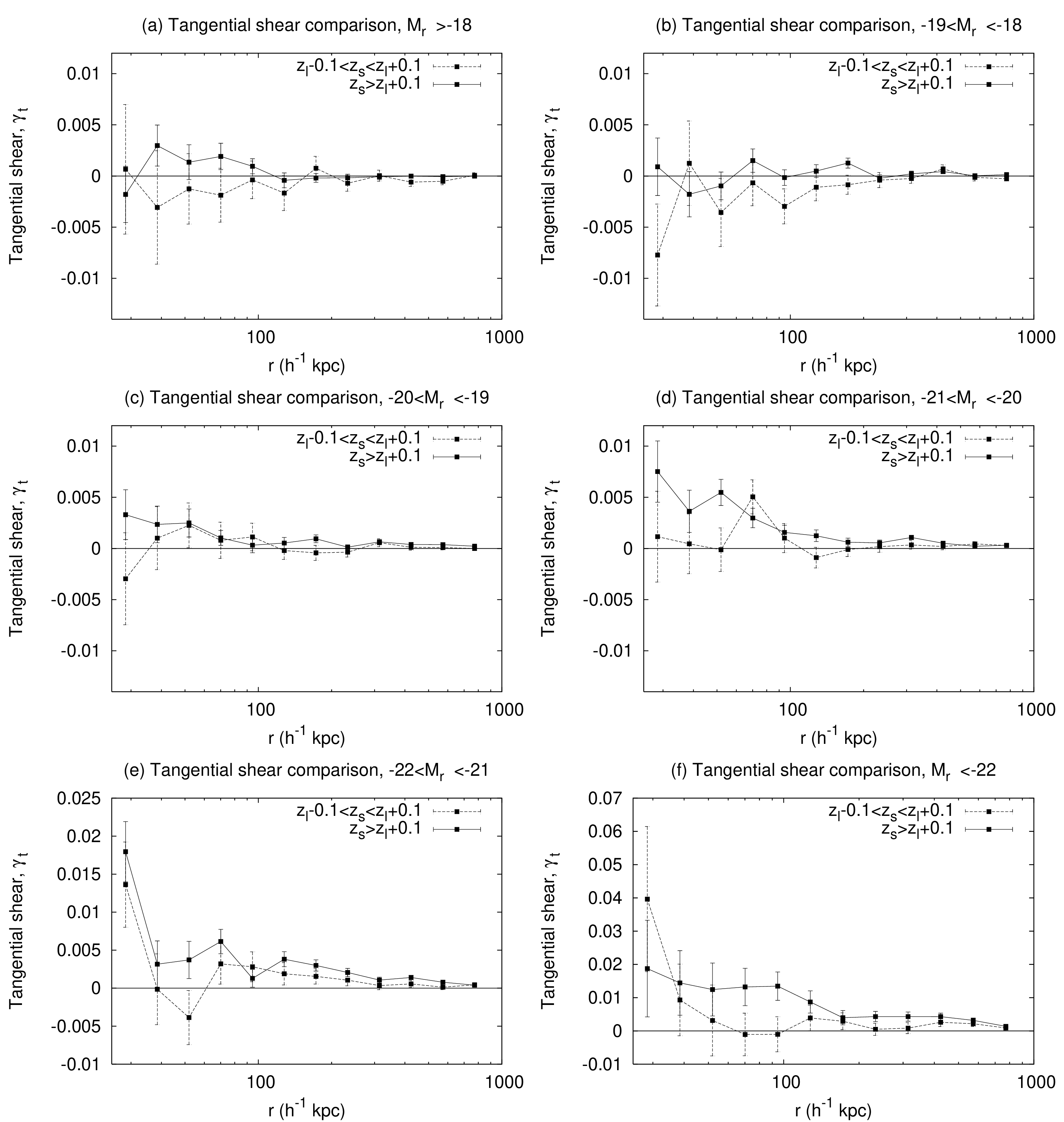}
\caption{\label{fig:emagranges}Shear as a function of radial pair
separation; panels (a)--(f) show the six luminosity bins in order of 
increasing luminosity.  The solid line shows the widely-separated 
lens-source pairs ($z_s-z_l>0.1$) and the dashed line shows the nearby 
pairs ($|z_s-z_l|<0.1$).  Each point on the plot shows three of our 
radial bins averaged together to increase the signal/noise ratio for 
visual inspection.  Note that the vertical scales are increased for panels 
(e) and (f).}
\end{figure*}

\begin{table*} 
\caption{\label{tab:shearlimit}Limits on the tangential intrinsic shear
$\Delta\gamma$, for various subsamples and ranges of radii (statistical
errors only).  The $10^3\Delta\gamma$ column shows the central (most
favored) value.  Quantities have been corrected to account for dilution by
non-physically associated galaxies in the $|z_s-z_l|<\epsilon$ samples as
described in the text.  The statistical uncertainties ($1\sigma$) on
$\Delta\gamma$ are shown in the next two columns for the
random-catalog and bootstrap methods. $\Delta\gamma_{gl}$ is the estimated
value of shear for these samples assuming a simple model for contamination
by pairs that are actually gravitationally lensed (the errors are
$1\sigma$ statistical errors on shear measurement only, i.e. they do not
include uncertainties associated with the source redshift distribution).
The next two columns show the pair ratios $R_p$ for the
$|z_s-z_l|<\epsilon$ (``in'') and $z_s>z_l+\epsilon$ (``out'') samples.  
Finally, we show confidence intervals for $\Delta\gamma$. The ``95\%
stat.'' confidence interval is computed using the central value of
$\Delta\gamma-\Delta\gamma_{gl}$ and statistical errors only.  The
``99.9\% stat.+sys.'' includes identified systematics as discussed in the
text.  In the row containing the $[-\infty,+\infty]$ confidence interval,
we find no detection of physically associated lens-source pairs when the
allowed range of systematic errors is taken into account.}
\begin{tabular}{ccrrrrrrrrcrr} \hline\hline
Subsample & $r$ & $10^3\Delta\gamma$ & 
\multicolumn{2}{c}{\mbox{$10^3\sigma(\Delta\gamma)$}}
& $10^3\Delta\gamma_{gl}$ & $R_p$[in]
& $R_p$[out] & 
\multicolumn{5}{c}{\mbox{$10^3\Delta\gamma$ confidence interval}} \\
 & & & rand & boot & & & & \multicolumn{2}{c}{\mbox{95\% stat.}}
& & \multicolumn{2}{c}{\mbox{99.9\% stat.+sys.}} \\
 & & & & & & & & min & max & & min & max \\
\hline
\multicolumn{13}{c}{\mbox{$\epsilon=0.1$}} \\
\hline
all	& $[	30	,	446	]$ & $	1.6	$ & $	1.0	$ 
& $	1.0	$ & $	1.45	\pm	0.17	$ & $	1.155	$ & $	
1.008	$ & $	-1.8	$ & $	2.1	$ & & $	-6.2	$ & $	6.6	$ 
\\
all	& $[	30	,	100	]$ & $	2.7	$ & $	2.1	$ 
& $	1.7	$ & $	1.96	\pm	0.32	$ & $	1.355	$ & $	
1.007	$ & $	-3.4	$ & $	5.0	$ & & $	-10.5	$ & $	12.5	$ 
\\
all	& $[	100	,	300	]$ & $	-0.4	$ & $	1.1	$ 
& $	1.3	$ & $	1.26	\pm	0.22	$ & $	1.181	$ & $	
1.010	$ & $	-4.1	$ & $	0.8	$ & & $	-9.5	$ & $	5.1	$ 
\\
all	& $[	300	,	446	]$ & $	3.6	$ & $	1.7	$ 
& $	1.6	$ & $	1.58	\pm	0.25	$ & $	1.120	$ & $	
1.007	$ & $	-1.4	$ & $	5.4	$ & & $	-7.4	$ & $	12.8	$ 
\\
$M_{r}>-18$	& $[	30	,	446	]$ & $	-19.6	$ & $	
14.9	$ & $	17.7	$ & $	-0.19	\pm	3.05	$ & $	1.022	$ 
& $	0.991	$ & $	-54.2	$ & $	15.3	$ & & $	-612.4	$ & $	
305.5	$ \\
$M_{r}>-18$	& $[	30	,	100	]$ & $	-18.1	$ & $	
25.4	$ & $	22.9	$ & $	9.16	\pm	4.20	$ & $	1.068	$ 
& $	0.990	$ & $	-77.7	$ & $	23.3	$ & & $	-206.6	$ & $	
115.3	$ \\
$M_{r}>-18$	& $[	100	,	300	]$ & $	-24.0	$ & $	
24.6	$ & $	21.4	$ & $	-3.13	\pm	4.34	$ & $	1.024	$ 
& $	0.991	$ & $	-69.8	$ & $	28.1	$ & & $	-412.1	$ & $	
243.1	$ \\
$M_{r}>-18$	& $[	300	,	446	]$ & $	-15.5	$ & $	
24.0	$ & $	31.4	$ & $	-0.25	\pm	5.02	$ & $	1.017	$ 
& $	0.990	$ & $	-76.7	$ & $	46.3	$ & & $	-\infty	$ & $	
\infty	$ \\
$-19$ to $-18$	& $[	27	,	493	]$ & $	-0.7	$ & $	
3.3	$ & $	3.7	$ & $	1.52	\pm	0.59	$ & $	1.079	$ 
& $	1.013	$ & $	-9.5	$ & $	5.1	$ & & $	-28.6	$ & $	
20.8	$ \\
$-19$ to $-18$	& $[	27	,	100	]$ & $	-17.5	$ & $	
9.5	$ & $	10.0	$ & $	0.11	\pm	1.56	$ & $	1.148	$ 
& $	1.010	$ & $	-37.3	$ & $	2.1	$ & & $	-70.2	$ & $	
21.5	$ \\
$-19$ to $-18$	& $[	100	,	300	]$ & $	-6.9	$ & $	
5.4	$ & $	5.2	$ & $	1.26	\pm	0.88	$ & $	1.094	$ 
& $	1.014	$ & $	-18.8	$ & $	2.5	$ & & $	-47.6	$ & $	
19.6	$ \\
$-19$ to $-18$	& $[	300	,	493	]$ & $	6.1	$ & $	
5.0	$ & $	5.3	$ & $	1.89	\pm	0.86	$ & $	1.068	$ 
& $	1.013	$ & $	-6.1	$ & $	14.6	$ & & $	-29.0	$ & $	
45.3	$ \\
$-20$ to $-19$	& $[	33	,	545	]$ & $	1.1	$ & $	
1.8	$ & $	1.6	$ & $	1.53	\pm	0.33	$ & $	1.119	$ 
& $	1.004	$ & $	-4.1	$ & $	3.3	$ & & $	-10.8	$ & $	
9.7	$ \\
$-20$ to $-19$	& $[	33	,	100	]$ & $	6.2	$ & $	
5.2	$ & $	4.8	$ & $	1.56	\pm	0.78	$ & $	1.248	$ 
& $	1.002	$ & $	-5.8	$ & $	15.1	$ & & $	-18.4	$ & $	
29.8	$ \\
$-20$ to $-19$	& $[	100	,	300	]$ & $	-0.7	$ & $	
2.6	$ & $	2.4	$ & $	1.28	\pm	0.49	$ & $	1.149	$ 
& $	1.007	$ & $	-7.1	$ & $	3.1	$ & & $	-15.9	$ & $	
10.7	$ \\
$-20$ to $-19$	& $[	300	,	545	]$ & $	1.6	$ & $	
2.4	$ & $	2.3	$ & $	1.67	\pm	0.45	$ & $	1.102	$ 
& $	1.004	$ & $	-4.8	$ & $	4.7	$ & & $	-12.9	$ & $	
12.8	$ \\
$-21$ to $-20$	& $[	27	,	735	]$ & $	2.6	$ & $	
1.3	$ & $	1.3	$ & $	1.18	\pm	0.20	$ & $	1.139	$ 
& $	1.011	$ & $	-1.1	$ & $	3.9	$ & & $	-5.8	$ & $	
9.8	$ \\
$-21$ to $-20$	& $[	27	,	100	]$ & $	6.6	$ & $	
3.5	$ & $	2.6	$ & $	2.59	\pm	0.57	$ & $	1.427	$ 
& $	1.007	$ & $	-3.0	$ & $	11.0	$ & & $	-13.2	$ & $	
23.2	$ \\
$-21$ to $-20$	& $[	100	,	300	]$ & $	-1.0	$ & $	
2.0	$ & $	1.8	$ & $	1.32	\pm	0.28	$ & $	1.217	$ 
& $	1.014	$ & $	-6.2	$ & $	1.6	$ & & $	-14.1	$ & $	
7.8	$ \\
$-21$ to $-20$	& $[	300	,	735	]$ & $	3.5	$ & $	
1.5	$ & $	1.6	$ & $	1.08	\pm	0.26	$ & $	1.119	$ 
& $	1.010	$ & $	-0.8	$ & $	5.6	$ & & $	-5.8	$ & $	
12.7	$ \\
$-22$ to $-21$	& $[	33	,	992	]$ & $	2.8	$ & $	
1.1	$ & $	1.2	$ & $	1.57	\pm	0.21	$ & $	1.146	$ 
& $	1.006	$ & $	-1.1	$ & $	3.6	$ & & $	-5.7	$ & $	
8.8	$ \\
$-22$ to $-21$	& $[	33	,	100	]$ & $	3.6	$ & $	
3.3	$ & $	2.3	$ & $	1.80	\pm	0.54	$ & $	1.807	$ 
& $	1.027	$ & $	-4.8	$ & $	8.5	$ & & $	-14.4	$ & $	
19.0	$ \\
$-22$ to $-21$	& $[	100	,	300	]$ & $	3.3	$ & $	
1.8	$ & $	1.8	$ & $	2.00	\pm	0.37	$ & $	1.355	$ 
& $	1.019	$ & $	-2.2	$ & $	4.8	$ & & $	-8.9	$ & $	
12.4	$ \\
$-22$ to $-21$	& $[	300	,	992	]$ & $	2.6	$ & $	
1.4	$ & $	1.6	$ & $	1.48	\pm	0.26	$ & $	1.121	$ 
& $	1.005	$ & $	-2.0	$ & $	4.1	$ & & $	-7.2	$ & $	
10.1	$ \\
$M_{r}<-22$	& $[	45	,	992	]$ & $	5.3	$ & $	
1.6	$ & $	1.5	$ & $	2.25	\pm	0.37	$ & $	1.309	$ 
& $	1.048	$ & $	-0.2	$ & $	6.4	$ & & $	-7.8	$ & $	
18.0	$ \\
$M_{r}<-22$	& $[	45	,	100	]$ & $	-0.6	$ & $	
6.7	$ & $	3.9	$ & $	4.59	\pm	1.28	$ & $	2.818	$ 
& $	1.204	$ & $	-18.4	$ & $	8.0	$ & & $	-43.7	$ & $	
29.8	$ \\
$M_{r}<-22$	& $[	100	,	300	]$ & $	3.5	$ & $	
2.8	$ & $	2.2	$ & $	2.29	\pm	0.53	$ & $	1.855	$ 
& $	1.101	$ & $	-4.4	$ & $	6.9	$ & & $	-15.3	$ & $	
18.8	$ \\
$M_{r}<-22$	& $[	300	,	992	]$ & $	6.3	$ & $	
1.9	$ & $	1.8	$ & $	2.28	\pm	0.44	$ & $	1.246	$ 
& $	1.042	$ & $	0.2	$ & $	7.8	$ & & $	-8.2	$ & $	
22.1	$ \\
\hline \multicolumn{13}{c}{\mbox{$\epsilon=0.05$}} \\ \hline
all	& $[	30	,	446	]$ & $	-0.1	$ & $	0.9	$ 
& $	1.2	$ & $	0.61	\pm	0.08	$ & $	1.201	$ & $	
1.014	$ & $	-2.9	$ & $	1.6	$ & & $	-7.1	$ & $	5.1	$ 
\\
all	& $[	30	,	100	]$ & $	-0.7	$ & $	2.4	$ 
& $	1.8	$ & $	0.97	\pm	0.15	$ & $	1.458	$ & $	
1.022	$ & $	-6.4	$ & $	3.0	$ & & $	-14.1	$ & $	9.7	$ 
\\
all	& $[	100	,	300	]$ & $	-1.3	$ & $	1.2	$ 
& $	1.4	$ & $	0.50	\pm	0.10	$ & $	1.234	$ & $	
1.017	$ & $	-4.6	$ & $	0.9	$ & & $	-9.6	$ & $	4.5	$ 
\\
all	& $[	300	,	446	]$ & $	1.5	$ & $	1.6	$ 
& $	1.9	$ & $	0.68	\pm	0.11	$ & $	1.155	$ & $	
1.012	$ & $	-2.9	$ & $	4.4	$ & & $	-8.5	$ & $	10.7	$ 
\\
$M_{r}>-19$	& $[	30	,	446	]$ & $	-4.5	$ & $	
4.1	$ & $	4.8	$ & $	0.51	\pm	0.32	$ & $	1.067	$ 
& $	1.004	$ & $	-14.5	$ & $	4.4	$ & & $	-29.9	$ & $	
15.8	$ \\
$M_{r}>-19$	& $[	30	,	100	]$ & $	-12.2	$ & $	
10.6	$ & $	9.0	$ & $	0.87	\pm	0.65	$ & $	1.135	$ 
& $	1.005	$ & $	-34.2	$ & $	8.0	$ & & $	-65.4	$ & $	
31.2	$ \\
$M_{r}>-19$	& $[	100	,	300	]$ & $	-8.4	$ & $	
5.5	$ & $	5.7	$ & $	0.00	\pm	0.45	$ & $	1.076	$ 
& $	1.006	$ & $	-19.6	$ & $	2.8	$ & & $	-38.5	$ & $	
14.7	$ \\
$M_{r}>-19$	& $[	300	,	446	]$ & $	1.0	$ & $	
6.8	$ & $	7.8	$ & $	0.96	\pm	0.54	$ & $	1.055	$ 
& $	1.004	$ & $	-15.2	$ & $	15.3	$ & & $	-36.9	$ & $	
36.9	$ \\
$-21$ to $-19$	& $[	25	,	545	]$ & $	-0.4	$ & $	
1.2	$ & $	1.1	$ & $	0.58	\pm	0.08	$ & $	1.196	$ 
& $	1.015	$ & $	-3.3	$ & $	1.4	$ & & $	-7.7	$ & $	
5.1	$ \\
$-21$ to $-19$	& $[	25	,	100	]$ & $	-0.8	$ & $	
3.0	$ & $	2.6	$ & $	1.14	\pm	0.21	$ & $	1.469	$ 
& $	1.018	$ & $	-7.8	$ & $	4.0	$ & & $	-17.0	$ & $	
12.2	$ \\
$-21$ to $-19$	& $[	100	,	300	]$ & $	-2.4	$ & $	
1.5	$ & $	1.6	$ & $	0.50	\pm	0.12	$ & $	1.249	$ 
& $	1.018	$ & $	-6.0	$ & $	0.3	$ & & $	-12.0	$ & $	
4.0	$ \\
$-21$ to $-19$	& $[	300	,	545	]$ & $	0.9	$ & $	
1.7	$ & $	1.7	$ & $	0.57	\pm	0.11	$ & $	1.163	$ 
& $	1.013	$ & $	-3.1	$ & $	3.7	$ & & $	-8.7	$ & $	
9.6	$ \\
$M_{r}<-21$	& $[	33	,	992	]$ & $	2.3	$ & $	
1.0	$ & $	1.0	$ & $	0.61	\pm	0.07	$ & $	1.253	$ 
& $	1.018	$ & $	-0.4	$ & $	3.7	$ & & $	-3.4	$ & $	
8.0	$ \\
$M_{r}<-21$	& $[	33	,	100	]$ & $	2.0	$ & $	
3.8	$ & $	2.1	$ & $	1.00	\pm	0.24	$ & $	2.365	$ 
& $	1.100	$ & $	-6.5	$ & $	8.5	$ & & $	-16.8	$ & $	
19.5	$ \\
$M_{r}<-21$	& $[	100	,	300	]$ & $	2.8	$ & $	
1.7	$ & $	1.4	$ & $	0.85	\pm	0.13	$ & $	1.614	$ 
& $	1.049	$ & $	-1.4	$ & $	5.3	$ & & $	-6.3	$ & $	
11.6	$ \\
$M_{r}<-21$	& $[	300	,	992	]$ & $	2.1	$ & $	
1.3	$ & $	1.2	$ & $	0.58	\pm	0.08	$ & $	1.209	$ 
& $	1.015	$ & $	-1.0	$ & $	4.1	$ & & $	-4.7	$ & $	
9.0	$ \\
\hline\hline
\end{tabular}
\end{table*}

\subsection{Intrinsic shear}
\label{ss:ic}

Table~\ref{tab:shearlimit} shows the limits on intrinsic shear that can be
computed for the various luminosity subsamples and sets of physical radii.  
The shear was computed by averaging the shear over the range of radii
listed, weighting by the total number of lens-source pairs in each radial bin,
\begin{eqnarray}
w_i \!\! &=& \!\! {N(r_i) \over \sum_i N(r_i)},
\nonumber \\
\Delta\gamma \!\! &=& \!\! {N\over N-N_{rand}} \sum_i w_i \gamma_t(r_i)
= {R_p\over R_p-1} \sum_i w_i \gamma_t(r_i),
\label{eq:widg}
\end{eqnarray}
where $N$ and $N_{rand}$ are summed over all relevant radii to account for
the dilution by non-physically associated pairs.  The factor of
$N/(N-N_{rand}) = R_p/(R_p-1)$ is used because intrinsic shear only
applies to the excess pairs $N(r_i)-N_{rand}(r_i)$ (see
Appendix~\ref{app:dgdef}).  This number depends on
the set of radii and luminosities considered, and as can be seen from
Fig.~\ref{fig:excess.mag}, is large for fainter foregrounds and for
larger values of radii.  (It ranges from close to 50 for the faintest
subsamples, down to 1.8 for the brightest subsamples and smallest radii.)  
Note that shears and errors in all figures have {\it not} been corrected
to account for this dilution effect, and it has no effect on the $\chi^2$
values in table~\ref{tab:chi2} because the dilution factor cancels out.  

The errors on $\Delta\gamma$ were computed using both random catalogs
and bootstrap (see Sec.~\ref{sec:uncertainty}). First, the averaged shear
was computed for each of 78 random catalog outputs (using the same
weighting), and $\sigma^{(rand)}$ is the standard deviation of these
averaged shears.  This random catalog error was then used to compute the
confidence intervals using the Student's $t$-distribution, which is quite
close to Gaussian for 77 degrees of freedom.  Second, the error from the
bootstrap covariance matrices was estimated as
\begin{equation}
\sigma^{(boot)} = \sqrt{\sum_{i,j} w_i w_j C_{boot,ij}}
\end{equation}
and confidence intervals were computed using a Gaussian distribution
(since the noise in the bootstrap covariance matrix ${\mathbfss C}_{boot}$
is unknown).  In computing the confidence intervals in
Table~\ref{tab:shearlimit}, we used the greater of the two errors
(bootstrap or random catalog).

As a test of consistency, we compared the errors $\sigma^{(rand)}$ and
$\sigma^{(boot)}$ computed from the random catalogs and from the bootstrap
covariance matrices.  For most of the samples shown in the table, the
difference between the two error values was less than 25 per cent; the
exceptions are the $r<100 h^{-1}$~kpc bins for the brighter lens samples,
for which the bootstrap errors are smaller than the random catalog errors.  
This is probably due to the fact that the number density of source
galaxies in the $|z_s-z_l|<\epsilon$ samples is greater near lens galaxies
than in the field, hence the shape noise is suppressed.  This effect is
taken into account by the bootstrap but not by the random catalogs, hence
the bootstrap errors are smaller.

A possible source of contamination to the intrinsic alignment signal is
gravitational lensing of source galaxies at $z_s>z_l$. We have attempted
to estimate the magnitude of such contamination ($10^3\Delta\gamma_{gl}$
in the table) according to the following procedure.  First, the mean value
of the inverse critical surface density $\Sigma_c^{-1}(z_s,z_l)$ for each
sample was computed, where $\Sigma_c^{-1}=0$ for $z_s<z_l$.  Then, the
averaged shear value $\gamma$ for the radii of interest was computed.  
The estimated contamination is then
\begin{eqnarray}
\Delta\gamma_{gl} &=&
\frac{\Sigma_c^{-1}[|z_s-z_l|<\epsilon]}{\Sigma_c^{-1}[z_s>z_l+\epsilon]} 
\left(\frac{N}{N-N_{rand}}\right) \gamma[z_s>z_l+\epsilon],
\nonumber \\ && \;
\end{eqnarray}
where $N/(N-N_{rand})$ is the dilution factor mentioned previously for the
$|z_s-z_l|<\epsilon$ samples.  The value $\Delta\gamma_{gl}$ should thus
be subtracted from $\Delta\gamma$ in order to yield the correct
intrinsic shear. While this calculation gives an estimate of the shear
signal due to contamination by lensed pairs, it does not take into account
the (currently unknown) photometric redshift error distribution.  In
particular, $\Sigma_c^{-1}[|z_s-z_l|<\epsilon]$ is close to zero because
the lens-source separation is small, and hence its fractional systematic
error may be large (possibly of order unity).

The central values of the confidence intervals in
Table~\ref{tab:shearlimit} labeled ``95\% stat.'' are computed by
subtracting the estimated gravitational lensing contamination from the
observed intrinsic alignment signal
($\Delta\gamma-\Delta\gamma_{gl}$). The statistical error on
$\Delta\gamma_{gl}$ is negligible and has not been included.

The ``99.9\% stat.+sys.'' confidence interval includes, in addition to
statistical errors, three of the possible systematic errors: calibration
biases in the shear measurement (Sec.~\ref{ss:calib}), errors in the
``correction factor'' $R_p/(R_p-1)$ of Eq.~(\ref{eq:widg}) caused by
seeing-induced correlations between high- and low-redshift galaxies (see
Sec.~\ref{ss:pzt}), and systematic errors in $\Delta\gamma_{gl}$ (the
statistical errors are negligible).  The former is taken into account by
dividing the computed shear by a worst-case calibration factor of 0.82
(appropriate if $\delta\gamma/\gamma=-0.18$).  The latter is taken into
account by assigning to $R_p$ a possible systematic error equal to
$2|R_p[{\rm out}]-1|$, where $R_p[{\rm out}]$ is the pair ratio for the
$z_s>z_l+\epsilon$ sample.  This is appropriate since $R_p[{\rm out}]=1$
in the absence of magnification bias, photometric redshift errors, and
seeing-induced systematics; it may even be a conservative estimate of
seeing-induced systematics because some of the $R_p[{\rm out}]>1$ signal
observed for brighter subsamples is likely due to photometric redshift
errors.  Since the weakest constraints are always obtained from the
smallest $R_p$, we replace $R_p$ in Eq.~(\ref{eq:widg}) with
$R_p-2|R_p[{\rm out}]-1|$. (In the outermost radial bin for the $\mrp>-18$
luminosity subsample, this ``minimum value'' of $R_p$ is less than one,
i.e. when systematic errors are taken into account we do not have a firm
detection of clustering of source galaxies around this sample of lenses.)
Finally, we allow $\Delta\gamma_{gl}$ to range from zero to twice the
measured value, i.e. the ``maximum'' value of the confidence interval for
$\Delta\gamma$ is computed assuming no gravitational lensing contamination
at all, and the ``minimum'' value is computed by subtracting
$2\Delta\gamma_{gl}$ instead of only $\Delta\gamma_{gl}$ from
$\Delta\hat\gamma$.

As can be seen from Table~\ref{tab:shearlimit}, we find no evidence for 
intrinsic alignment of the satellite galaxies (there is a ``detection'' 
at 95 per cent confidence for the $\mrp<-22$ subsample at 
$45<r<100h^{-1}$~kpc, but given the large number of subsamples 
investigated this should not be taken as evidence for $\Delta\gamma\neq 
0$).  With all samples averaged together, we find conservatively that 
$-0.0062<\Delta\gamma<+0.0066$, although we cannot exclude the possibility 
of a large intrinsic alignment if present around only the faintest 
``lens'' galaxies.

\subsection{Systematic shear}
\label{ss:icsys}

\begin{figure}
\includegraphics[width=3in]{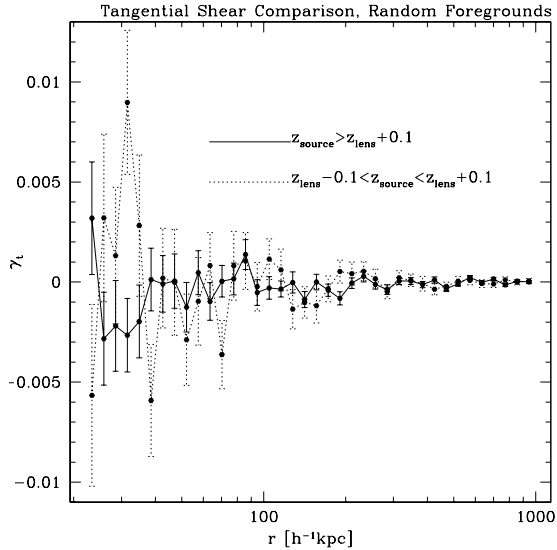}
\caption{\label{fig:shearrand}The tangential shear signal computed around 
one of the random lens catalogs.}
\end{figure}

Table~\ref{tab:sysshearlimit} includes limits on the systematic shear,
i.e. the additive bias contributed by observational systematics to the
total shear. Such a contribution could arise if, e.g. there were a
correlation between the PSF and the survey boundaries.  These limits were
computed by taking the average value of the shears computed from the 78
random catalogs, and finding confidence intervals around this average
using the Student's $t$-distribution with 77 degrees of freedom (dividing
$\sigma$ by $\sqrt{78}$ because the measurement was done with 78 random
catalogs).  As shown, the systematic shear is consistent with zero for all
samples; furthermore it is negligible compared to the statistical error in
$\Delta\gamma$, as can be seen by comparing the upper limits in
Table~\ref{tab:sysshearlimit} to the statistical errors shown in
Table~\ref{tab:shearlimit}.  A plot of the systematic shear is shown in 
Fig.~\ref{fig:shearrand}.

\begin{table}
\caption{\label{tab:sysshearlimit}Limits on systematic shear in the source
catalog $10^{5}\Delta\gamma_S$ for various subsamples and ranges of radii.  
Values are quoted using the errors computed from random catalogs ($rand$),
at 95 and 99.9 per cent confidence level, using the Student's
$t$-distribution (with errors divided by $\sqrt{78}$ because there are 78
measurements of systematic shear).}
\begin{tabular}{ccrr}
\hline\hline
Subsample & $r$ & \multicolumn{2}{c}{\mbox{$10^5\Delta\gamma_S^{(rand)}$}} 
\\
$\mrp$ & $h^{-1}$ kpc & 95\% & 99.9\% \\
\hline
\multicolumn{4}{c}{\mbox{$\epsilon=0.1$}} \\
\hline
all & $[30, 446]$ & $[-3.1,  2.7]$ & $[-5.2,  4.8]$ \\
all & $[30, 100]$ & $[-4.6, 20.2]$ & $[-13.5, 29.1]$ \\
all & $[100, 300]$ & $[-5.6,  1.8]$ & $[-8.2,  4.4]$ \\
all & $[300, 446]$ & $[-3.8,  4.4]$ & $[-6.8,  7.4]$ \\
$>-18$ & $[30, 446]$ & $[-1.5, 13.1]$ & $[-6.7, 18.3]$ \\
$>-18$ & $[30, 100]$ & $[-13.3, 59.9]$ & $[-39.5, 86.1]$ \\
$>-18$ & $[100, 300]$ & $[-5.2, 21.2]$ & $[-14.6, 30.6]$ \\
$>-18$ & $[300, 446]$ & $[-6.2, 11.4]$ & $[-12.6, 17.8]$ \\
$-19$ to $-18$ & $[27, 493]$ & $[-6.9,  3.9]$ &
$[-10.8,  7.8]$ \\
$-19$ to $-18$ & $[27, 100]$ & $[-14.3, 40.9]$ &
$[-34.1, 60.7]$ \\
$-19$ to $-18$ & $[100, 300]$ & $[-9.5, 11.3]$ &
$[-16.9, 18.7]$ \\
$-19$ to $-18$ & $[300, 493]$ & $[-11.0,  3.4]$ &
$[-16.1,  8.5]$ \\
$-20$ to $-19$ & $[33, 545]$ & $[-5.8,  3.0]$ &
$[-9.0,  6.2]$ \\
$-20$ to $-19$ & $[33, 100]$ & $[-21.7, 25.3]$ &
$[-38.5, 42.1]$ \\
$-20$ to $-19$ & $[100, 300]$ & $[-10.0,  5.0]$ &
$[-15.3, 10.3]$ \\
$-20$ to $-19$ & $[300, 545]$ & $[-6.1,  3.9]$ &
$[-9.7,  7.5]$ \\
$-21$ to $-20$ & $[27, 735]$ & $[-4.6,  2.4]$ &
$[-7.1,  4.9]$ \\
$-21$ to $-20$ & $[27, 100]$ & $[-17.7, 29.7]$ &
$[-34.7, 46.7]$ \\
$-21$ to $-20$ & $[100, 300]$ & $[-11.9,  3.9]$ &
$[-17.5,  9.5]$ \\
$-21$ to $-20$ & $[300, 735]$ & $[-4.3,  2.9]$ &
$[-6.9,  5.5]$ \\
$-22$ to $-21$ & $[33, 992]$ & $[-2.4,  3.6]$ &
$[-4.6,  5.8]$ \\
$-22$ to $-21$ & $[33, 100]$ & $[-32.3, 34.7]$ &
$[-56.4, 58.8]$ \\
$-22$ to $-21$ & $[100, 300]$ & $[-20.3,  0.7]$ &
$[-27.8,  8.2]$ \\
$-22$ to $-21$ & $[300, 992]$ & $[-1.6,  5.0]$ &
$[-4.1,  7.5]$ \\
$<-22$ & $[45, 992]$ & $[-9.7,  7.7]$ & $[-16.0, 14.0]$ \\
$<-22$ & $[45, 100]$ & $\!\![-120.7, 72.7]$ & $[-190.1, 142.1]$ \\
$<-22$ & $[100, 300]$ & $[-32.7, 26.3]$ & $[-53.9, 47.5]$ \\
$<-22$ & $[300, 992]$ & $[-8.6,  8.2]$ & $[-14.7, 14.3]$ \\
\hline
\multicolumn{4}{c}{\mbox{$\epsilon=0.05$}} \\
\hline
all & $[30, 446]$ & $[-4.1,  2.7]$ & $[-6.6,  5.2]$ \\
all & $[30, 100]$ & $[-5.4, 28.2]$ & $[-17.4, 40.2]$ \\
all & $[100, 300]$ & $[-5.9,  3.9]$ & $[-9.4,  7.4]$ \\
all & $[300, 446]$ & $[-6.6,  3.2]$ & $[-10.1,  6.7]$ \\
$>-19$ & $[30, 446]$ & $[-3.7,  8.1]$ & $[-7.9, 12.3]$ \\
$>-19$ & $[30, 100]$ & $[-11.6, 45.0]$ & $[-31.9, 65.3]$ \\
$>-19$ & $[100, 300]$ & $[-2.9, 14.7]$ & $[-9.2, 21.0]$ \\
$>-19$ & $[300, 446]$ & $[-9.9,  6.3]$ & $[-15.6, 12.0]$ \\
$\-21$ to $-19$ & $[25, 545]$ & $[-7.8,  0.8]$ & $[-10.9,  
3.9]$ \\
$\-21$ to $-19$ & $[25, 100]$ & $[-1.0, 41.8]$ &
$[-16.4, 57.2]$ \\
$\-21$ to $-19$ & $[100, 300]$ & $[-11.6,  2.0]$ & 
$[-16.6,  7.0]$ \\
$\-21$ to $-19$ & $[300, 545]$ & $[-9.7,  1.1]$ &  
$[-13.6,  5.0]$ \\
$<-21$ & $[33, 992]$ & $[-5.1,  4.1]$ & $[-8.5,  7.5]$ \\
$<-21$ & $[33, 100]$ & $[-74.1, 24.1]$ & 
$\!\![-109.3, 59.3]$ \\
$<-21$ & $[100, 300]$ & $[-18.6, 10.0]$ & $[-28.9, 20.3]$ \\
$<-21$ & $[300, 992]$ & $[-4.6,  5.4]$ & $[-8.2,  9.0]$ \\
\hline
\hline
\end{tabular}
\end{table}

\section{Discussion}
\label{sec:discussion}

In this paper, we have set stringent new limits on intrinsic alignments as
a contaminant of the galaxy-galaxy lensing signal.  For example, averaging
together all lens galaxies at radii out to 446$h^{-1}$~kpc we derive
$-0.0062<\Delta\gamma<+0.0066$ (99.9 per cent confidence including
identified systematics; see Table~\ref{tab:shearlimit}) or
$-0.0018<\Delta\gamma<+0.0021$ (95 per cent confidence, statistical errors
only).  The statistical constraint is a factor of $\sim 5$ ($\sim 1.5$ for
our more conservative error bar) tighter than $|\Delta\gamma|\le 0.01$ (95
per cent confidence) as obtained from APM/2dF data by
\citet{2002AJ....124..733B}.  It also extends to smaller radii than the
measurements of nearby spirals by \citet{2001ApJ...555..106L}.  
Furthermore, our study of the luminosity and radius dependence of the
intrinsic shear is useful for weak lensing studies that divide lens
galaxies into luminosity subsamples.

Our limits have been obtained by using ``lens'' galaxies with
spectroscopic redshifts and ``source'' galaxies with photometric
redshifts, a different strategy than that selected by previous observers
(\citealt{2001ApJ...555..106L}; \citealt{2002AJ....124..733B}) who used
lens-source pairs selected entirely on the basis of spectroscopic redshift
information. This strategy has dramatically increased our sample size and
reduced our statistical uncertainties for the brighter lenses ($\mrp<-20$)
because these are typically at redshift $z>0.1$, where most of their
satellites are too faint to be targeted by SDSS spectroscopy.  This method
does however have disadvantages. For our fainter-luminosity lenses, the
photometric-redshift source method gives very large error bars
(Table~\ref{tab:shearlimit}) because there are few physically associated
lens-source pairs and hence any intrinsic alignment signal will be drowned
out by the large number of unassociated pairs.  The use of photometrically
selected pairs also comes with the caveat that the limits on
$\Delta\gamma$ could be circumvented by a small class of galaxies that
have strong intrinsic alignments but always produce incorrect photometric
redshifts.  It is thus recommended that the values in
Table~\ref{tab:shearlimit} with redshift slice half-width $\epsilon=0.1$
be used to constrain intrinsic alignments in galaxy-galaxy lensing
studies, since in this case only about 20 per cent of the physically
associated source galaxies are scattered out of the sample by photometric
redshift errors.

Despite the recent interest in intrinsic alignments as a contaminant of
cosmic shear (i.e. shear-shear correlations), there has been comparatively
little theoretical work on the density-shear correlation.
\citet{2001ApJ...555..106L} do provide a prediction for the density-shape
correlation for spiral ``source'' galaxies, which we convert (see
Appendix~\ref{app:pac}) to $\Delta\gamma=-0.004$ at small radii
$r<1h^{-1}$~Mpc.  This is marginally consistent with our results (it is
within our more conservative confidence interval of
$-0.0062<\Delta\gamma<+0.0066$), although the comparison is imperfect as
the model applies only to spiral sources.  More generally, measurements of
the density-shear correlation $\Delta\gamma$ provide a constraint on
intrinsic alignment models in addition to that provided by the shear
autocorrelation measurements (\citealt{2000ApJ...543L.107P};
\citealt{2002MNRAS.333..501B}).  It would thus be useful to compute the
predictions of other intrinsic alignment models for $\Delta\gamma$ and
determine which models agree with the intrinsic shear measurements
presented here and in \citet{2002AJ....124..733B}.

We conclude by roughly estimating the intrinsic alignment contamination to
the gravitational lensing shear, $(1-R_p^{-1})\Delta\gamma$.  For example,
if our source catalog restricted to $z_s>z_l+0.1$ were used in a lensing
study of the halos of SDSS spectroscopic galaxies, we would have a
contamination fraction $1-R_p^{-1}\sim 0.01$ in the $100<r<300 h^{-1}$~kpc
bin and intrinsic alignment contamination $|\Delta\gamma|<0.0095$ (see
Table~\ref{tab:shearlimit}), corresponding to a contamination to the
gravitational lensing shear at the level of $<9.5\times 10^{-5}$.  This is
approximately equal to the size of the $1\sigma$ error bars at these radii
for our full sample ($1.0\times 10^{-4}$; see Fig.~\ref{fig:e.and.bmode})  
and a factor of six less than the shear signal, $\sim 5.8\times 10^{-4}$.  
Slightly weaker constraints are placed on the intrinsic alignments around
the brighter lenses, but in these cases the lensing signal is a factor of
several higher, e.g. for the $-21>M_{r}>-22$ magnitude range we have
$|\Delta\gamma|<0.0124$ at $100<r<300h^{-1}$~kpc, $1-R_p^{-1}\sim 0.02$,
and shear signal $\gamma_t= 2\times 10^{-3}$, so the contamination is at
worst about a factor of eight smaller than the signal. These results
indicate at 99.9 per cent confidence that, except possibly for the
faintest lens galaxies, intrinsic alignments are only contaminating the
sub-Mpc scale galaxy-galaxy lensing signal at the $\la 15$ per cent level.

\section*{Acknowledgments}

We acknowledge useful discussions with James Gunn, \v{Z}eljko Ivezi\'{c},
Yeong-Shang Loh, Robert Lupton, Patrick McDonald, and Michael Strauss.

C.H. is supported by a NASA Graduate Student Researchers Program
(GSRP) fellowship.  R.M. is supported by an NSF Graduate Research 
Fellowship.  U.S. is supported by Packard Foundation, Sloan Foundation,
NASA NAG5-1993, and NSF CAREER-0132953.

Funding for the creation and distribution of the SDSS Archive has been
provided by the Alfred P. Sloan Foundation, the Participating
Institutions, the National Aeronautics and Space Administration, the
National Science Foundation, the U.S. Department of Energy, the Japanese
Monbukagakusho, and the Max Planck Society. The SDSS Web site is {\slshape
http://www.sdss.org/}.

The SDSS is managed by the Astrophysical Research Consortium (ARC) for the
Participating Institutions. The Participating Institutions are The
University of Chicago, Fermilab, the Institute for Advanced Study, the
Japan Participation Group, The Johns Hopkins University, Los Alamos
National Laboratory, the Max-Planck-Institute for Astronomy (MPIA), the
Max-Planck-Institute for Astrophysics (MPA), New Mexico State University,
University of Pittsburgh, Princeton University, the United States Naval
Observatory, and the University of Washington.

\appendix

\section{Intrinsic shear and correlation function}
\label{app:dgdef}

In principle, the tangential component of intrinsic shear $\Delta\gamma_t$ 
is a function not only of the transverse separation $r$ between the lens 
and source galaxy, but also on the line-of-sight separation $u$.  The 
contribution of the intrinsic alignments to the shear $\hat\gamma_t$ 
observed in galaxy-galaxy weak lensing is:
\begin{equation}
\hat\gamma_t({\rm intr},r) = \frac{ \int \Delta\gamma_t(r,u) n(r,u) 
\rmd u }{ \int n(r,u) \rmd u },
\label{eq:d1d}
\end{equation}
where $n(r,u)\rmd u\rmd^2{\bmath r}$ is the number of sources with 
line-of-sight separation between $u$ and $u+\rmd u$ in area 
$\rmd^2{\bmath r}$ in the lens plane.  This contamination is typically 
estimated by comparing the total number of sources observed per unit area 
in the lens plane, i.e. $\int n(r,u) \rmd u$, to the number of sources 
observed in a randomly selected field, $\int n(\infty, u)\rmd u$.  We 
express this in terms of a ``pair ratio'' $R_p(r)$:
\begin{equation}
R_p(r) = \frac{ \int n(r,u) \rmd u }{ \int n(\infty, u)\rmd u }.
\label{eq:d2d}
\end{equation}
The simplest interpretation of the pair ratio is that the fraction of the 
source galaxies that are physically unassociated with the lens is 
$R_p^{-1}$, and that the remaining fraction $1-R_p^{-1}$ are physically 
associated with the lens.  This motivates the definition of the 
projected intrinsic shear $\Delta\gamma(r)$ of physically associated 
galaxies as:
\begin{equation}
\Delta\gamma(r) \equiv \frac{\hat\gamma_t({\rm intr},r)}{1-R_p^{-1}}.
\label{eq:dgdef}
\end{equation}
Most analyses of the effect of intrinsic alignments on galaxy-galaxy
lensing studies (e.g. \citealt{2002AJ....124..733B};  
\citealt{2003astro.ph.12036S}) have used Eq.~(\ref{eq:dgdef}) to compute
$\hat\gamma_t({\rm intr})$ from the measured pair ratio $R_p$ and an
estimate of the intrinsic shear of physically associated galaxies
$\Delta\gamma(r)$ based on spectroscopically selected lens-source pairs at
the same redshift.

While Eq.~(\ref{eq:dgdef}) is directly useful for measuring the intrinsic 
alignment and estimating the contamination of weak lensing results, we can 
also write $\Delta\gamma(r)$ in terms of a projection integral along the 
line of sight.  Substituting Eqs.~(\ref{eq:d1d}) and (\ref{eq:d2d}) into
Eq.~(\ref{eq:dgdef}), we find:
\begin{equation}
\Delta\gamma(r) = \frac{\int \Delta\gamma_t(r,u) n(r,u) \rmd u}
{\int [n(r,u)-n(\infty,u)] \rmd u}.
\label{eq:dgn}
\end{equation}
The integrands in both the numerator and denominator of Eq.~(\ref{eq:dgn}) 
vanish at large $|u|$ (so long as the magnification bias due to 
gravitational lensing is small).  At small $|u|$ (i.e. much less than the 
typical distance to the lenses, so that variations in $n(\infty,u)$ can be 
ignored), $n(r,u)$ is related to the galaxy-galaxy correlation function 
via $n(r,u)\propto 1+\xi(\sqrt{r^2+u^2})$.  Therefore we may write
\begin{equation}
\Delta\gamma(r)
= \frac{ \int \Delta\gamma_t(r,u) [ 1 + \xi(\sqrt{r^2+u^2}) ]\rmd u}
{\int \xi(\sqrt{r^2+u^2})\rmd u},
\label{eq:dgxi}
\end{equation}
which is the relation of $\Delta\gamma(r)$ to fundamental (as 
opposed to observed) quantities.

\section{Relation of intrinsic shear to position angle statistics}
\label{app:pac}

\citet{2000ApJ...532L...5L} and \citet{2001ApJ...555..106L} measure
density-shape correlations using the position angle statistic,
\begin{equation}
\omega_{2D}(r_{3D}) \equiv \langle \cos^2\phi\rangle - {1\over 2}
= {1\over 2} \langle \cos 2\phi\rangle,
\label{eq:omega2d}
\end{equation}
where $\phi$ is the position angle of the source galaxy's minor axis
relative to the lens ($0$ indicates radial alignment, $\pi/2$ tangential
alignment) and the three-dimensional pair separation is
$r_{3D}=\sqrt{r^2+u^2}$.  \citet{2001ApJ...555..106L} consider spiral
galaxies and hence interpret the minor axis as the (projected) rotation
axis of the galaxy, but Eq.~(\ref{eq:omega2d}) is what is directly
measured.

We consider here a crude model for the conversion between $\omega_{2D}$ 
and the intrinsic shear $\Delta\gamma$.  Formulas for the transformation 
of ellipticity under shear are given by \citet{1991ApJ...380....1M}; using 
that $\tan 2\phi=e_{45}/e_t$ (in a coordinate system aligned with the 
direction to the lens galaxy) we can find the transformation of the 
position angle under shear,
\begin{equation}
{\partial\phi\over \partial\gamma_t} = -e^{-1}\sin 2\phi.
\end{equation}
Substitution into Eq.~(\ref{eq:omega2d}) yields
\begin{equation}
{d\omega_{2D}\over d\gamma_t} = {1\over 2}\left\langle \left(
   -e^{-1}\sin 2\phi \right){d\over d\phi} \cos 2\phi \right\rangle
  = {1\over 2}\langle e^{-1}\rangle.
\label{eq:o2g}
\end{equation}
We define this quantity to be the position angle responsivity ${\cal
R}_\omega$; for our $r<19$ galaxies (with negligible measurement noise in
the ellipticity), $\langle e^{-1}\rangle =2.75$ and ${\cal
R}_\omega=1.38$.  The theory of \citet{2001ApJ...555..106L} suggests that
at small radii ($r_{3D}<1.5 h^{-1}$~Mpc) we should have $\omega_{2D}(r) =
-0.0057$ for spirals (see their Fig.~2) or $\gamma_t = \omega_{2D}/{\cal
R}_\omega = -0.0041$.  Since the galaxy autocorrelation function
$\xi(r_{3D})\gg 1$ at these radii, we expect $\Delta\gamma=-0.0041$ (cf
Eq.~\ref{eq:dgxi}).  This conversion is only a rough estimate since the
relation Eq.~(\ref{eq:o2g}) between $\omega_{2D}$ and $\gamma_+$ was
derived for a gravitational shear, and it need not be the correct relation
for intrinsic shear.

Using the conversion ${\cal R}_\omega=1.38$, the data presented in Fig.~2 
of \citet{2001ApJ...555..106L} correspond to $\Delta\gamma = -0.0037\pm 
0.0025$ at $r_{3D}=500h^{-1}$~kpc.

\section{Noise-rectification bias}
\label{app:kn}

Noise in galaxy images has several effects for weak lensing.  The
best-known effect is the introduction of measurement noise in the
ellipticity.  It has also been pointed out that noise can also couple to
the PSF anisotropy to produce spurious power in cosmic shear measurements,
even if the PSF correction is otherwise exact 
(\citealt{2000ApJ...537..555K}; \citealt{2002AJ....123..583B}).  This 
occurs because of a noise-rectification bias, i.e. the ellipticity is a 
nonlinear function of the image intensity so that an unbiased estimate of 
the intensity need not translate into an unbiased estimate of the 
ellipticity.  Here we estimate the effect of noise-rectification biases on 
the shear calibration of low signal-to-noise ratio galaxies.

Our plan for approaching this problem is as follows:
\begin{enumerate}
\item Determine, for a circular Gaussian galaxy with unit radius 
$\sigma=1$ and unit central intensity $I(0)=1$, the mean and covariance of 
the measured adaptive moments $\hat {\mathbfss M}^{(I)}$ to leading order 
in the noise.
\item Use scaling and symmetry principles to derive the mean and 
covariance of $\hat{\mathbfss M}^{(I)}$ for a general elliptical Gaussian 
galaxy.
\item Propagate these errors to the PSF-corrected ellipticity 
$\hat{\mathbfss e}^{(f)}$.
\end{enumerate}
Here we will neglect corrections due to non-Gaussianity of the galaxy or 
PSF.

We first introduce some notation.  We write the normalization, centroid
and moments of the galaxy as a six-dimensional vector:
\begin{equation}
c^\alpha = (A,x_0,y_0,[M^{-1}]^{xx},[M^{-1}]^{xy},[M^{-1}]^{yy}).
\label{eq:calpha}
\end{equation}
(We could have used the matrix elements of ${\mathbfss M}$ to 
parameterize the family of Gaussians instead of ${\mathbfss M}^{-1}$; 
however using ${\mathbfss M}^{-1}$ simplifies the algebra.)  We now 
introduce the notation: \begin{equation}
J({\bmath r}; c^\alpha) = A \exp -{({\bmath r}-{\bmath r}_0)^T
{\mathbfss M}^{-1} ({\bmath r}-{\bmath r}_0)\over 2}.
\label{eq:j}
\end{equation}
Thus the energy functional Eq.~(\ref{eq:energy}) becomes
\begin{equation}
E(c^\alpha; I) = {1\over 2}\int [ I({\bmath r}) - J({\bmath r}, c^\alpha) 
]^2 \rmd^2{\bmath r};
\end{equation}
or, written as an inner product:
\begin{equation}
E(c^\alpha; I) = {1\over 2}\left< I-J, I-J\right>.
\end{equation}
(We suppress the $c^\alpha$ argument of $J$ for clarity.)  Minimizing this 
yields:
\begin{equation}
0 = \left< I-J, J_\alpha\right>,
\label{eq:eminb}
\end{equation}
where $J_\alpha\equiv\partial J/\partial c^\alpha$.

We may obtain the dependence of $c^\alpha$ on $I$ by taking the functional
derivative of Eq.~(\ref{eq:eminb}) with respect to $I({\bmath r}_1)$:
\begin{equation}
0=J_\alpha({\bmath r}_1) + \left( 
\left< I-J, J_{\alpha\beta} \right> -\left< J_\beta, J_\alpha \right>
\right) {\rmd c^\beta\over \delta I({\bmath r}_1)}.
\label{eq:ed1}
\end{equation}
The second derivative of Eq.~(\ref{eq:eminb}) is more complicated; for the 
particular case $I=J$, it is:
\begin{eqnarray}
0 &=& {\rmd^2c^\beta\over \delta I({\bmath r}_1) \delta I({\bmath r}_2)}
\left( \left< J_{\alpha\beta}, I-J\right> - \left< J_\beta, J_\alpha 
\right> \right)
\nonumber \\ &&
- {\rmd c^\beta \over\delta I({\bmath r}_1)}
  {\rmd c^\gamma\over\delta I({\bmath r}_2)}
\bigl( \left< J_{\alpha\beta}, J_\gamma \right>
     + \left< J_{\beta\gamma}, J_\alpha \right>
\nonumber \\ && \; \; \; \;
     + \left< J_{\gamma\alpha}, J_\beta \right>
     - \left< J_{\alpha\beta\gamma}, I-J \right> \bigr)
\nonumber \\ &&
+ {\rmd c^\beta \over\delta I({\bmath r}_1)} J_{\alpha\beta}({\bmath r}_2)
+ {\rmd c^\beta \over\delta I({\bmath r}_2)} J_{\alpha\beta}({\bmath r}_1).
\label{eq:ed2}
\end{eqnarray}
Defining ${\mathbfss H}$ to be the matrix inverse of $[{\mathbfss
H}^{-1}]_{\alpha\beta} = \left< J_\alpha, J_\beta\right>$, and evaluating 
the derivatives at $I=J$, gives
\begin{equation}
\left.{\rmd c^\beta \over\delta I({\bmath r})} \right|_{I=J} = 
H^{\alpha\beta} J_\alpha({\bmath r})
\label{eq:deriv1}
\end{equation}
and
\begin{equation}
\int
\left.{\rmd^2c^\beta\over \delta I({\bmath r}) \delta I({\bmath r})}
\right|_{I=J} \rmd^2{\bmath r} =
- H^{\alpha\beta} H^{\gamma\delta} 
\left< J_\alpha, J_{\gamma\delta} \right>.
\label{eq:deriv2}
\end{equation}
In Eq.~(\ref{eq:deriv2}) we have only computed the trace of the 
second derivative matrix; this is all we will need.

A circular Gaussian galaxy with unit radius and central intensity has 
measured profile:
\begin{equation}
\hat I({\bmath r}) = \rme^{-(x^2+y^2)/2} + \eta({\bmath r})
= J({\bmath r}; C^\alpha) + \eta({\bmath r}),
\end{equation}
where $\eta$ is the noise and $C^\alpha=(1,0,0,1,0,1)$ are the moments of 
the galaxy (in the absence of noise).  The noise is assumed to be white:
\begin{equation}
\langle \eta({\bmath r}_1)\eta({\bmath r}_2)\rangle = 
{\cal N}\delta^{(2)}({\bmath r}_1 - {\bmath r}_2).
\label{eq:noisepower}
\end{equation}
The mean and covariance of the measured moments $\hat 
c^\alpha$ can then be determined to order $O({\cal N})$ from the 
equations:
\begin{equation}
\langle\hat c^\alpha \rangle = C^\alpha + {1\over 2}{\cal N}\int
\left.{\rmd^2c^\alpha\over \delta I({\bmath r}) \delta I({\bmath r})}
\right|_{I=J} \rmd^2{\bmath r},
\end{equation}
and:
\begin{equation}
{\rm Cov}(\hat c^\alpha, \hat c^\beta) = {\cal N} \int \left.
{\rmd c^\alpha\over I({\bmath r})}
{\rmd c^\beta \over I({\bmath r})}\right|_{I=J} \rmd^2{\bmath r}.
\end{equation}
Substituting Eqs.~(\ref{eq:deriv1}) and (\ref{eq:deriv2}) yields:
\begin{equation}
\langle \hat c^\alpha \rangle = C^\alpha - {1\over 2}{\cal N}
H^{\alpha\beta} H^{\gamma\delta}
\int J_\beta({\bmath r}) J_{\gamma\delta}({\bmath r}) \rmd^2{\bmath r}
\label{eq:mean-c}
\end{equation}
and
\begin{equation}
{\rm Cov}(\hat c^\alpha, \hat c^\beta) = {\cal N} H^{\alpha\beta}.
\label{eq:cov-c}
\end{equation}

The substitution of the particular functional form Eq.~(\ref{eq:j}) into 
Eqs.~(\ref{eq:mean-c}) and (\ref{eq:cov-c}) is a straightforward but 
lengthy exercise.  The ${\mathbfss H}$ matrix in the basis of 
Eq.~(\ref{eq:calpha}) is
\begin{equation}
{\mathbfss H} = {2\over\pi} \left( \begin{array}{cccccc}
 1 &  0 &  0 & 1 &  0 & 1 \\
 0 &  1 &  0 & 0 &  0 & 0 \\
 0 &  0 &  1 & 0 &  0 & 0 \\
 1 &  0 &  0 & 4 &  0 & 0 \\
 0 &  0 &  0 & 0 &  2 & 0 \\
 1 &  0 &  0 & 0 &  0 & 4
\end{array} \right),
\end{equation}
and the product $H^{\gamma\delta}J_{\gamma\delta}({\bmath r})$ appearing 
in Eq.~(\ref{eq:mean-c}) is
\begin{equation}
H^{\gamma\delta}J_{\gamma\delta}({\bmath r}) =
{2\over\pi}( \rho^4 -2) \rme^{-\rho^2/2}.
\end{equation}
From this we find that the mean values of the entries in ${\mathbfss 
M}^{-1}$ are
\begin{equation}
\langle [\hat M^{-1}]^{xx} \rangle = \langle [\hat M^{-1}]^{yy} \rangle 
= 1+{8\over\pi}{\cal N},
\end{equation}
and $\langle [\hat M^{-1}]^{xy} \rangle = 0$.  Using the Taylor series for 
the moments:
\begin{equation}
\hat{\mathbfss M} = {\mathbfss 1} - (\hat{\mathbfss M}^{-1}-{\mathbfss 1})
+ (\hat{\mathbfss M}^{-1}-{\mathbfss 1})^2 + O(\hat{\mathbfss 
M}^{-1}-{\mathbfss 1})^3,
\end{equation}
we can find the mean of $\hat{\mathbfss M}$ to be
\begin{equation}
\langle \hat M_{ij} \rangle = \left(1+{4\over\pi}{\cal N}\right)
\delta_{ij},
\label{eq:d1}
\end{equation}
and the covariance to be:
\begin{equation}
{\rm Cov}(\hat M_{ij},\hat M_{kl}) = {4\over\pi}{\cal N} (
\delta_{ik}\delta_{jl} + \delta_{il}\delta_{jk}).
\label{eq:d2}
\end{equation}

We now generalize Eqs.~(\ref{eq:d1}) and (\ref{eq:d2}).  Clearly the 
energy functional is translation-invariant, so 
Eqs.~(\ref{eq:d1},\ref{eq:d2}) cannot depend on ${\bmath r}_0$.  If the 
amplitude $A$ is increased by some factor, and the noise amplitude 
$\sqrt{\cal N}$ is similarly increased, the energy minimization remains 
unchanged, so the result can depend on $A$ and ${\cal N}$ only through the 
ratio $A/\sqrt{\cal N}$.  The problem is also invariant under linear 
transformations of ${\bmath r}$, so that if ${\mathbfss M}$ is changed, 
and ${\cal N}$ is increased by the factor $\sqrt{\det{\mathbfss M}}$ (to 
account for the delta function in Eq.~\ref{eq:noisepower} and the change 
of measure in the integration over $\rmd^2{\bmath r}$), then the linear 
transformation acts simply on Eqs.~(\ref{eq:d1},\ref{eq:d2}).  We have 
concluded that the generalizations of Eqs.~(\ref{eq:d1},\ref{eq:d2}) are
\begin{equation}
\langle \hat M_{ij} \rangle =  \left(1
+4{{\cal N}\over \pi A^2\sqrt{\det{\mathbfss M}}}\right) M_{ij}
\label{eq:dm1}
\end{equation}
and
\begin{equation}
{\rm Cov}(\hat M_{ij},\hat M_{kl}) = 4{{\cal N}\over
\pi A^2\sqrt{\det{\mathbfss M}}}(M_{ik}M_{jl} + M_{il}M_{jk}).
\label{eq:dm2}
\end{equation}
The quantity ${\cal N}/\pi A^2\sqrt{\det{\mathbfss M}}$ is recognized 
as $\nu^{-2}$, where $\nu$ is the signal-to-noise ratio for detection of 
the galaxy in an adaptive elliptical Gaussian filter 
\citep{2002AJ....123..583B}.  

We have reached the final stage in our plan, namely to translate the 
errors in ${\mathbfss M}$ into the calibration bias of the ellipticity 
estimator.  We consider here only the simplest case, that of a circular 
Gaussian PSF and an elliptical Gaussian galaxy with its major axis aligned 
along the $x$-axis.  The estimated ellipticity is:
\begin{equation}
\hat e_+^{(f)} = {\hat Q^{(I)} \over \hat T^{(I)}-T^{(P)}},
\label{eq:ehatf}
\end{equation}
where $T$ is the trace and $Q=M_{xx}-M_{yy}$ is the $+$ quadrupole moment.  
The effects of the bias (Eq.~\ref{eq:dm1}) and noise (Eq.~\ref{eq:dm2}) 
in ${\mathbfss M}^{(I)}$ can be determined from the first and second 
derivatives (respectively) of $\hat e_+^{(f)}$ with respect to ${\mathbfss 
M}^{(I)}$; the order $O({\cal N})$ terms are:
\begin{eqnarray}
\langle \hat e_+^{(f)} \rangle &=&
e_+^{(f)} + {\langle\delta \hat Q^{(I)}\rangle \over T^{(I)}-T^{(P)}}
- {Q^{(I)} \langle \delta \hat T^{(I)}\rangle \over (T^{(I)}-T^{(P)})^2}
\nonumber \\ &&
- {{\rm Cov}(\hat Q^{(I)}, \hat T^{(I)})\over (T^{(I)}-T^{(P)})^2}
+ {Q^{(I)} \langle (\delta\hat T^{(I)})^2 \rangle \over 
(T^{(I)}-T^{(P)})^3},
\end{eqnarray}
where we have used the bias $\langle \delta\hat Q\rangle = \langle \hat 
Q\rangle-Q$.  For a galaxy with only $+$ ellipticity (i.e. position 
angle along the $x$- or $y$-axis), the biases and covariances obtained 
from Eqs.~(\ref{eq:dm1},\ref{eq:dm2}) are:
\begin{eqnarray}
\langle \delta \hat Q^{(I)} \rangle   &=& 4\nu^{-2}T^{(I)}e_+^{(I)};
\nonumber \\
\langle \delta \hat T^{(I)} \rangle   &=& 4\nu^{-2}T^{(I)};
\nonumber \\
\langle(\delta \hat T^{(I)})^2\rangle &=& 4\nu^{-2}T^{(I)2}(1+e_+^{(I)2});
\nonumber \\
{\rm Cov}(\hat Q^{(I)}, \hat T^{(I)}) &=& 8\nu^{-2}T^{(I)2}e_+^{(I)}.
\nonumber \\
\langle(\delta \hat Q^{(I)})^2\rangle &=& 4\nu^{-2}T^{(I)2}(1+e_+^{(I)2});
\label{eq:dqt}
\end{eqnarray}
This leads us to the result for the change in ellipticity calibration:
\begin{equation}
{\langle\delta\hat e_+^{(f)}\rangle \over e_+^{(f)} } = 
{4\over\nu^2}(1-3R_2^{-1}+R_2^{-2}+e_+^{(f)2}).
\label{eq:noiserect}
\end{equation}
We have verified Eqs.~(\ref{eq:dm1}), (\ref{eq:dm2}), and
(\ref{eq:noiserect}) in ``toy'' simulations.

We have used this result as an approximation to the shear calibration 
$\delta\gamma/\gamma$, although this is not exact because the shear 
calibration is determined by the derivative of $\langle\delta\hat 
e_+^{(f)}\rangle$ with respect to $e_+^{(f)}$, which is not exactly equal 
to the ratio $\langle\delta\hat e_+^{(f)}\rangle/e_+^{(f)}$ when the 
latter is not constant.  The calibration bias obtained from 
Eq.~(\ref{eq:noiserect}) is shown in Fig.~\ref{fig:knp}.

\section{Chi-squared test with errors from simulations}
\label{app:chi2}

We perform several tests for consistency with zero signal using a $\chi^2$ 
test.  Since we have derived our covariance matrix from simulations, we 
must also take into account the noise from only having a finite number of 
simulations.

If we have estimated a shear in $N$ radial bins, we can construct an 
$N$-dimensional shear vector ${\bmath\gamma}$.  The $M$ simulations 
provide us with $M$ simulated shear vectors 
$\{{\bmath\gamma}^{(\alpha)}\}_{\alpha=1}^M$ from which we can compute the 
sample covariance $\hat{\mathbfss C}$ (see Eq.~\ref{eq:sample-cov}).  We 
are interested in the distribution of the variable
\begin{equation}
\hat\chi^2 = {\bmath\gamma}^T\hat{\mathbfss C}^{-1}{\bmath\gamma}
\end{equation}
under the null hypothesis that ${\bmath\gamma}$ is Gaussian with mean 
${\bmath 0}$ and the same covariance ${\mathbfss C}$ as the simulated 
shear vectors.

In order to determine the distribution of $\hat\chi^2$ via Monte Carlo, we 
define the vectors ${\bmath x}={\mathbfss C}^{-1/2}{\bmath\gamma}$ and 
${\bmath x}^{(\alpha)}={\mathbfss C}^{-1/2}{\bmath\gamma}^{(\alpha)}$.  
(Here the symmetric matrix ${\mathbfss C}^{-1/2}$ has the same 
eigenvectors as ${\mathbfss C}$ but the eigenvalues $\lambda_i$ are 
replaced by $\lambda_i^{-1/2}$.)  Then we find:
\begin{equation}
\hat\chi^2 = {\bmath x}^T {\mathbfss D}^{-1} {\bmath x},
\end{equation}
where the matrix ${\mathbfss D}$ is given by:
\begin{equation}
{\mathbfss D} = {1\over M-1} \left[ \sum_{\alpha=1}^M
{\bmath x}^{(\alpha)} {\bmath x}^{(\alpha)T} -
\bar{\bmath x}\bar{\bmath x}^T \right].
\label{eq:x}
\end{equation}
Here $\bar{\bmath x}$ are the sample means of the ${\bmath 
x}^{(\alpha)}$.  Since the $x_i$ and $x^{(\alpha)}_i$ 
are independent Gaussians of zero mean and unit variance,
it is computationally easy to generate sample from
Eq.~(\ref{eq:x}), so we can use a Monte Carlo method to compute the 
probability distribution for $\hat\chi^2$.

As $M\rightarrow\infty$ (with fixed $N$), ${\mathbfss D}\rightarrow
{\mathbfss 1}$ and the distribution of $\hat\chi^2$ converges to the
$\chi^2$ distribution. However, the convergence is not rapid.  To see
this, we note that by Wick's theorem, ${\mathbfss D}$ has expectation
value equal to the identity ${\mathbfss 1}$ and covariance:
\begin{equation}
\langle (D_{ij}-\delta_{ij})(D_{kl}-\delta_{kl}) \rangle
= {\delta_{ik}\delta_{jl}+\delta_{il}\delta_{jk}\over M-1}.
\end{equation}
Using the Taylor series for the inverse of a matrix, and the fact that the 
higher-order moments of the ${\mathbfss D}$ distribution decrease as 
$M^{-2}$ or faster, we find the asymptotic expansion:
\begin{equation}
\langle{\mathbfss D}^{-1}\rangle = {\mathbfss 1} +
\langle  ({\mathbfss D}-{\mathbfss 1})^2 \rangle + O(M^{-2}).
\end{equation}
Since ${\bmath x}$ is independent of ${\mathbfss D}$:
\begin{equation}
\langle \hat\chi^2 \rangle = N + {N(N+1)\over M} + O(M^{-2}),
\label{eq:nn1}
\end{equation}
which is found to agree with the mean obtained via Monte Carlo.  Since the 
standard deviation of the $\chi^2$ distribution is $\sqrt{2N}$, 
we must take $M\gg N^{3/2}$ in order for the $O(1/M)$ correction term in 
Eq.~(\ref{eq:nn1}) to be small compared to the width of the $\chi^2$ 
distribution.  Therefore the standard $\chi^2$ test is only valid if $M\gg 
N^{3/2}$ (not true for our random lens test, with $M=78$ and $N=39$).

\end{document}